\documentclass[lettersize,journal]{IEEEtran}
\usepackage{amsmath,amsfonts}
\usepackage{array}
\usepackage{textcomp}
\usepackage{stfloats}
\usepackage{url}
\usepackage{verbatim}
\usepackage{graphicx}
\usepackage{balance}
\IEEEoverridecommandlockouts
\usepackage{cite}
\usepackage{mathrsfs,amsmath,amssymb,amsfonts}
\usepackage{cases}
\usepackage{titlesec}
\usepackage{comment}
\usepackage{xcolor}
\usepackage{algorithm} 
\usepackage{algpseudocode}
\usepackage{multirow}
\usepackage{tablefootnote}
\usepackage{pifont}
\usepackage{tabularx}
\usepackage[normalem]{ulem}
\usepackage{subcaption}
\usepackage{multirow}

\def\BibTeX{{\rm B\kern-.05em{\sc i\kern-.025em b}\kern-.08em
    T\kern-.1667em\lower.7ex\hbox{E}\kern-.125emX}}

\begin{document}

\title{ DT-DDNN: A Physical Layer Security Attack Detector in 5G RF Domain for CAVs}

\author{Ghazal Asemian,~\IEEEmembership{Student Member,~IEEE,} Mohammadreza Amini,~\IEEEmembership{Senior Member,~IEEE,} Burak Kantarci, ~\IEEEmembership{Senior Member,~IEEE} and Melike Erol-Kantarci,~\IEEEmembership{Fellow,~IEEE}% <-this % stops a space
\thanks{The authors are with the School of Electrical Engineering and Computer Science, University of Ottawa, Ottawa, ON, Canada, (gasem093@uottawa.ca, mamini6@uottawa.ca, burak.kantarci@uottawa.ca, melike.erolkantarci@uottawa.ca)}% <-this % stops a space

\vspace{-10mm}}

\markboth{IEEE Transaction on Vehicular Technology}%
{Over-The-Air Double-Threshold Deep Learner for
Jamming Detection in 5G RF domain for CAVs}

\maketitle
\thispagestyle{plain}
\pagestyle{plain}
\pagenumbering{gobble}

\begin{abstract}
The Synchronization Signal Block (SSB) is a fundamental component of the 5G New Radio (NR) air interface, crucial for the initial access procedure of Connected and Automated Vehicles (CAVs) and serves several key purposes in the network’s operation. However, due to the predictable nature of SSB transmission, including the Primary and Secondary Synchronization Signals (PSS and SSS), jamming attacks are critical threats. Leveraging radio frequency (RF) domain knowledge, this work presents a novel deep learning-based technique focusing on SSB for detecting jammers in CAV networks without changing pre-existing infrastructure. By integrating a preprocessing block that extracts PSS correlation and energy per null resource elements (EPNRE) characteristics, our method distinguishes between normal and jammed received signals with high precision. Additionally, by incorporating Discrete Wavelet Transform (DWT), the efficacy of training and detection are optimized. A double-threshold double Deep Neural Network (DT-DDNN) is also introduced to the architecture complemented by a deep cascade learning model to increase the sensitivity of the model to variations of signal-to-jamming noise ratio (SJNR). Results show that the proposed method achieves 96.4\% detection rate in extra low jamming power, i.e., SJNR between 15 to 30 dB which outperforms the single threshold DNN design with 86.0\% detection rate and unprocessed IQ sample DNN design with 83.2\% detection rate. Ultimately, the performance of DT-DDNN is validated through the analysis of real 5G signals obtained from a practical testbed, demonstrating a strong alignment with the simulation results.\end{abstract}

\vspace{-3mm}
\section{Introduction}
%\input{Sections-short/introduction1} 
%-------------------What is 5G SSB----------------------------------
The evolution in transformative technologies such as Connected and Automated Vehicles (CAVs), the Internet of Things (IoT), and Edge Computing, necessitates the development of next-generation wireless networks to guarantee the Quality of Service (QoS) for communication processes \cite{hamidi20215g}.
Given the impact of security attacks on QoS, various defense strategies, particularly machine learning-based techniques, are proposed to enhance network robustness \cite{khan2019survey}\cite{zhang2019towards}\cite{park2021comprehensive}\cite{amini24}\cite{Lohan.2024}.
Moreover, proper operation of critical elements of the network, such as synchronization signal block (SSB), is crucial for maintaining device-network synchronization and ensuring service integrity \cite{tuninato2023comprehensive}. 
% cite WWRF work
The SSB functions as a reference signal that guarantees that the user equipment (UE) and the base station operate at the same time and frequency. 
5G SSBs must be robust and resilient to a variety of channel conditions to guarantee synchronization that is reliable despite the complexity of the environment. In addition, 5G enables the configuration of SSBs to be more adaptable to various deployment scenarios and facilitates more effective network management.

SSB is transmitted based on predetermined frequency and timing resources in 5G networks \cite{3gpp.38.211}. Thus, by identifying the subcarrier spacing and getting synchronized with the cell in the time domain, an attacker can extract the cell identity and as a result, locate and target Primary Synchronization Signal (PSS) or Secondary Synchronization Signal (SSS) which prevents UE from receiving critical signals required for synchronization \cite{flores2023implementation}, \cite{pirayesh2022jamming}. 
Detecting PSS and SSS results in the extraction of Physical Cell Identity (PCI) which can lead to a jamming attack on the Physical Broadcast Channel (PBCH) and prevent UEs from accessing necessary information and new connections to cells \cite{lichtman20185g}, \cite{wang2023detecting}, \cite{pirayesh2022jamming}. %Authors in \cite{arjoune2020smart} suggested using localization-based detection techniques as a solution, however, these techniques are not efficient in a mobile jammer scenario.
Furthermore, unlike encrypted user data, control signals such as PSS and SSS are transmitted unencrypted \cite{wang2023detecting}. Thus, SSB can potentially be jammed by anyone without the requirement of deciphering or authentication. 
Jamming attacks on SSB require less jamming power to disrupt the communication \cite{ludant2021sigunder} enabling the jammer to reduce the detection probability as most of the detection systems rely on identifying anomalous high-energy patterns. Meanwhile, it helps the attacker to perform the attack with simple and inexpensive equipment.
Traditional jamming detection methods rely on the received signal intensity, or the performance of the network \cite{arjoune2020real}. The signal level detection methods are incapable of maintaining the detection accuracy in high signal-to-jamming and noise ratio (SJNR) scenarios, and are mostly efficient in constant jamming cases \cite{ornek2022efficient}, \cite{arjoune2020smart}. The detection techniques based on the network metrics such as packet delivery ratio (PDR) or bit error rate (BER) may be unsuccessful when facing advanced jammers utilizing selective or intelligent techniques which may not have a significant impact on the error rate or performance metrics.

%-------------------SSB Jamming Detection---------------------------
To address the security requirements of wireless networks, it is crucial to investigate solutions that operate effectively within the limitations of current infrastructure. This research presents a jamming detection technique for 5G networks based on a deep learning approach operating solely in the radio frequency (RF) domain, employing characteristics of RF signals such as energy patterns and signal correlations without the need for decryption or deep integration into network layers. This approach not only simplifies implementation by avoiding infrastructural modifications but also enables real-time security monitoring required for applications such as CAVs. 
A key discriminator that sets our work apart from existing jamming detection algorithms is its focus on the analysis of 5G SSB using the design of a double threshold double Deep Neural Network model (DT-DDNN) to enable the system to detect both smart and barrage jammers effectively.
This expands the algorithm's applicability and impact by covering a wider range of jamming scenarios. A critical component of our detection system is the preprocessing block designed to extract features from the received signal that are indicative of jamming activities. This block focuses on the correlation of the PSS and the measurement of Energy per Null Resource Elements (EPNRE), which are significant indicators of potential jamming. The PSS correlation provides a robust method for detecting synchronization disruptions commonly resulting from jamming signals, while the EPNRE metric helps identify unusual energy patterns in the SSB subcarriers where no transmission should occur in a normal transmission scenario. As the deep learning algorithms require advanced computational resources, we incorporate the Discrete Wavelet Transform (DWT) into our design. The DWT magnifies the features that were extracted during the preprocessing phase, and provides a more detailed representation of the signal attributes. This not only improves the performance of the detection model but also optimizes the training procedure. A significant enhancement to our architecture is the integration of double threshold Deep Neural Network (DNN) which has been used to improve precision in the scenarios indicated by higher SJNR values in which the detector faces challenges to make a final decision. The additional DNN is supported by a deep cascade learning model to increase the sensitivity of the design in a high SJNR regime. Therefore, the main contributions of this study are summarized as follows. 
\begin{itemize}
    \item  Utilize over-the-air 5G signal features without the need for higher layer key performance indicators (KPIs) such as Block Error Rate (BLER), Bad Packet Ratio (BPR), throughput, and other metrics in higher layers of 5G protocol stack. This enables the jamming detection module to be implemented independently in 5G networks,.i.e., the proposed method does not need to be implemented on 5G network entities such as gNodeB (gNB), UE, or 5G core.
    \item Exploit salient features in the SSB that are relevant to jamming signal through preprocessing of the received 5G waveform. Particularly, PSS correlation, DWT, and energy per null subcarriers in SSB are employed. Furthermore, Log transformation is applied to adjust the dynamic range of the extracted energy so that the jamming signal can be better distinguished from environmental noise.
    \item Implement a double-threshold deep learning structure to improve the detection performance in a high SJNR regime and optimize their thresholds. In particular, a double deep learning structure is proposed in which the first DNN uses two thresholds at its output to best determine any uncertainties in the detection process. Exploiting deep cascaded structure, the second DNN decides on the observations with high uncertainty in which the UE experiences very low jamming power. %SJNR between 15 to 30 dB. 
\end{itemize}

%--------------------------Structure--------------------------------

The two thresholds at the output of the first DNN are optimized in a way that 100\% empirical detection rate for the two classes is achieved, and the threshold for the second DNN is chosen empirically based on the target false alarm probability.
Using the proposed techniques enables DT-DDNN to attain a 96.4\% detection rate in low jamming power conditions when SJNR is between 15 dB to 30 dB. The performance improvement is significant when compared to 86.0\% detection rate of the single threshold DNN approach and 83.2\% detection rate of the unprocessed IQ sample DNN method. Furthermore, a testbed is developed for experimental evaluation of the proposed approach which validates the performance of the system in real-world applications.

The rest of this work is organized as follows: Section \ref{relatedwork} discusses the existing works in jamming detection in 5G. Section \ref{sysModel} provides a brief introduction to the 5G SSB and formulates the problem. Section \ref{detector} includes detailed information on the jamming detector design and each component of the architecture. Section \ref{results} presents the results of the detection technique. Finally, conclusions are provided in section \ref{conclusion}.\vspace{-2mm}

\section{Related Work}
\label{relatedwork}
%----------------Jamming Detection techniques-----------------------
% intro for related works
\begin{table*}[htbp]
    \caption{Overview of physical layer attack detection techniques in literature and this work } \vspace{-2 mm}
    \label{table:detection-compare}
    \centering
    \begin{tabular}{|c|c|c|c|c|}
    \hline
    Ref. & Technical Approach / Network Model & Feature or Parameter & Attack Type & Testbed \\
    \hline 
        \hline
    \cite{zhang2020physical} & Analytical (optimal power allocation)/NOMA & Transmit Power & NU Eavesdropper & \ding{55}\\
    \hline
    \cite{nandan2021beamforming} & Analytical(ZFBF)/CRN &  Beamforming Vectors & Eavesdropper & \ding{55}\\
    \hline
    \cite{huang2020physical} & Analytical (SOP\tablefootnote{secrecy outage probability})/mmWave NOMA & Transmit Power & Eavesdropper & \ding{55}\\ 
    \hline
    \cite{hossain2022physical} & Analytical (average secrecy capacity, SOP) & RIS element phases & Eavesdropper & \ding{55}\\
    \hline
    \cite{eltayeb2017enhancing} & Analytical (optimal beamforming)/ mmWave V2V & CSI\tablefootnote{channel state information}, Transmit Power& Eavesdropper& \ding{55}\\
    \hline
    \cite{sharma2022mitigating} & ML(DRL)/5G Het-Nets &channel, beamforming vectors & constant, adaptive, reactive& \ding{55}\\
    \hline
    \cite{viana2023deep} & ML(CNN, LSTM)/5G UAV&SINR, RSSI & jamming & \ding{55}\\
    \hline
    \cite{wang2023detecting} & statistical(SPCA) /5G &PBCH data & PBCH Jamming & \ding{55} \\
\hline
    \cite{arjoune2020real} & ML(Hoeffding Decision Tree)/5G& Received Signal & Barrage Jammer&\ding{55} \\
       \hline
    \cite{ornek2022efficient}\cite{ornek2023securing}& Statistical (EVM)/5G & EVM & Tone, Chirp Jammer& \checkmark\\
    \hline
    \cite{chiarello2021jamming} & Statistical (GLRT)/5G & NA & Smart Jammer & \ding{55} \\
   
    \hline
    \cite{jere2023bayesian}& ML(Supervised Learning)/5G & Cross-layer KPI & Various Jammer Types & \checkmark \\
    \hline
    \cite{kouassi2023application} & ML(Ensemble Learning)/5G & Network Traffic & Constant,Random,Deceptive,Reactive & \ding{55}\\
     \hline
    \cite{hachimi2020multi} & ML(DL with kernelized SVM)/5G C-RAN &  Network Traffic & constant,random,deceptive,reactive & \ding{55}\\
    \hline
    \cite{wang2022anonymous} & ML(Supervised, Unsupervised Learning)/5G& Network Metrics & WiFi Interference, controlled jamming& \checkmark\\
    \hline
    \cite{feng2018machine} &  ML(KNN, Decision tree, Random Forest)/WiFi& Packet delivery rate, RSS & Constant, Random, Reactive& \ding{55}\\
    \hline
     \cite{Varotto2023} &  ML(AE)/5G& I-Q samples & Gaussian, Uniform& \checkmark\\
    \hline
    This work & ML(DT-DDNN) / 5G RF-Domain & PSS Corr, EPNRE & Smart SSB \& Barrage Jamemr & \checkmark\\
    \hline
    \end{tabular}
    \vspace{-4mm}
\end{table*}
The primary aim of physical layer security (PLS) is to improve wireless network security by leveraging the unique characteristics of the physical layer \cite{solaija2022towards}. Significant research is focused on identifying attacks on the physical layer of wireless networks.
% eavesdropping, analytical (optimization problem(inputs: power allocation, constraints: channel gains, total power, SNR threshold, noise power), NOMA
Regarding 5G Non-Orthogonal Multiple Access (NOMA) vulnerabilities, \cite{zhang2020physical} discusses the issue of near-end user (NU) eavesdropping activities. Solutions include modifying serial interference cancellation (SIC) process through cooperative jamming to develop optimal power allocation strategies thereby improving the secrecy rates while maintaining the data throughput of legitimate users within acceptable limits.
% MIMO-NOMA
Authors in \cite{nandan2021beamforming} analyze how resource reusability compromises physical layer security in multiple-input multiple-output NOMA (MIMO-NOMA) suggesting zero-forcing beamforming (ZFBF) with signal alignment and eigen beamforming method is used to maximize the signal-to-information-leakage-plus-noise ratio (SLNR) and therefore improve the physical layer security. 
% mmWave NOMA, 
A minimal angle-difference user pairing scheme in millimeter wave (mmWave) NOMA network along with two secrecy beamforming models taking advantage of the spatial correlation between the user and attacker is provided in \cite{huang2020physical} to increase the secrecy rate in response to an eavesdropper.
% UOWC
The efficacy of physical layer security of the integration of reconfigurable intelligent surface (RIS) and Radio Frequency-Underwater Optical Wireless Communication (UOWC) in a scenario including an eavesdropper is investigated in \cite{hossain2022physical}. 
% mmWave vehicular
The research in \cite{eltayeb2017enhancing} develops a multiantenna transmission and beamforming model to increases the secrecy rate of mmWave vehicular communication.
% write a sentence to related PLS to jamming
% 5G Het-Net, jamming, federated learning
Authors in \cite{sharma2022mitigating} present a novel anti-jamming technique based on federated deep reinforcement learning (DRL) which is based on a joint beamforming and power allocation optimization problem. 
% 5G UAV
The study in \cite{viana2023deep} focuses on attack identification in air-to-ground communication links using deep attention recognition (DAtR) which detects security attacks using a small deep network embedded in legitimate unmanned aerial vehicles (UAVs). The system employs two metrics of received signal strength indicator (RSSI) and signal to interference plus noise ratio (SINR) to identify attacks in different scenarios such as non-line-of-sight (NLoS), line-of-sight (LoS), or a combination of both.
% Smart Jammer, statistical(based on blank REs), 5G and beyond
In \cite{chiarello2021jamming} a jamming detection and defense strategy is proposed by implementing pseudo-random blanking of resource elements within orthogonal frequency-devision mutiplexing (OFDM) symbols based on statistical hypothesis testing. The work is focused on smart jammers which maximize their impacts by reducing spectral efficiency and BLER values, and minimize their detection probability. Through the detection algorithm, certain subcarriers are left blank across OFDM symbols in a pseudo-random fashion. Adopting a pseudo-random algorithm that determines the blanking pattern makes the system unpredictable and increases its robustness to jamming. The suggested approach incorporates the downlink data transmission system of 5G without necessitating adjustments to the current infrastructure. 
% tone and chirp jammer, threshold-based (based on EVM), 5G
The authors in \cite{ornek2022efficient} and \cite{ornek2023securing} present a novel approach that employs the sensitivity of Error Vector Magnitude (EVM) to detect the presence of tone and chirp jammers when the EVM value reaches a predetermined threshold. Besides the high sensitivity and minimal complexity, this approach offers spectral information about the jammer and affected frequency bands.
% different types of jammer, ML with network parameters, 5G
In \cite{jere2023bayesian}, an adaptable, multidimensional strategy is introduced for detecting and classifying jamming attacks, considering power levels and frequency band variations. The detection technique is based on supervised learning and receives metrics such as channel quality indicator (CQI), bit rates, packet rates, and power headroom.
% (constant, random, deceptive, reactive) jammer, XGBOOST-Ensemble learning (using network traffic), 5G
% deceptive jamming: mimics the signal of legit BS
The primary focus of \cite{kouassi2023application} is the utilization of ensemble learning and the XGBOOST-ensemble learning combination as a machine learning-based jamming detection in C-RAN. Authors use the WSN-DS database to assess the performance of various machine learning algorithms. The WSN-DS database includes 374,661 samples as chosen features that encapsulate the behavioral patterns of network traffic under normal and jammed conditions. Data preprocessing is performed to set up the dataset for analysis by separating the independent and dependent variables. The feature extraction process is carried out using the XGBOOST algorithm which emphasizes important patterns within the data.
% barrage jammer, Hoeffding decision tree (received signal), 5G
The study in \cite{arjoune2020real} emphasizes the need for real-time detection and mitigation techniques. 
Hence, the authors make a contribution by introducing a technique for real-time detection of jamming attacks, utilizing the Hoeffding decision tree. This machine learning methodology facilitates real-time processing and addresses limitations encountered in conventional decision tree models.
% (constant, random, deceptive, reactive) jammer, ML-intrusion detection (DL with kernelized SVM)(energy consumption, is CH, ADV CH send, ADV SCH send, data sent to BS(number of packets), Distance CH to BS, Data received, ADV CH receives, Join REQ receive, Time), 5G C-RAN
The work in \cite{hachimi2020multi} presents a multi-stage machine learning-based intrusion detection system (ML-IDS) specifically designed for 5G Cloud Radio Access Network (C-RAN) which is designed to detect constant, random, deceptive, and reactive jammers. A Multilayer Perceptron (MLP), a deep learning algorithm, and a Kernelized Support Vector Machine (KSCM) compose the proposed ML-IDS. 
% various types of jammers, supervised ML(downlink/uplink bitrate, downlink/uplink packet rate, downlink/uplink retransmission rate, PUSCH SNR, CQI, power headroom, energy per resource element, uplink path loss, downlink/uplink MCS, and the average turbo decoder rate) unsupervised (waveform), 5G
The integration of supervised and unsupervised learning methods for jamming detection is introduced in \cite{wang2022anonymous}. Known jamming attacks are detected using a supervised learning model while an unsupervised anomaly detection method using auto-encoders is applied for unknown jamming types. The learning-based model is executed based on network parameters such as bitrate, packet rate, retransmission rate, and CQI.
In \cite{wang2023detecting}, the focus is on intelligent PBCH jamming (PBCH-IJ) attack in 5G NR which disrupts the Master Information Block (MIB) decoding by applying a sniffing attack to extract PCI information. To identify anomalies that can detect the jamming attack, sparse principal component analysis (SPCA) is used in the design along with an adaptive detection threshold.
% This work is on Wifi
The study in \cite{feng2018machine} implements a machine learning-based detection technique using Network Simulator 3 (NS-3) simulator for constant, random, and reactive jammers. Received signal strength, carrier sense time, noise, and PDR are used as jamming detection metrics. The detection method is tested based on three different learning algorithms K nearest neighbor (KNN), decision tree, and random forest with the highest accuracy of 81\%.
This work distinguishes itself from the aforementioned research by focusing on smart SSB jamming detection in the RF domain using signal features of a 5G network.
In this work, the smart SSB jammer detects the synchronization signal information on the 5G resource grid (including time, frequency, and pattern) and disrupts the corresponding resource blocks, making its detection highly challenging \cite{asemian2024impactmobilitybeamsweeping}. The detection technique is a machine learning-based model using two DNN blocks one with a single threshold and the second one with double threshold characteristics to be able to detect the existence of the jammer in high SJNR conditions. To further enhance the performance of the DNN model deep cascade learning approach is adapted by the second DNN block. A summary of the current literature is provided in Table \ref{table:detection-compare}. \vspace{-3mm}

\section{System Model and Problem Formulation}
\label{sysModel}
% Received SSB = y
%------------------------Background--------------------------------
\subsection{Background and Feasibility Analysis}
\begin{figure*}
    \centering
    \begin{subfigure}{0.35\textwidth}
        \includegraphics[width=1\textwidth, height=4cm]{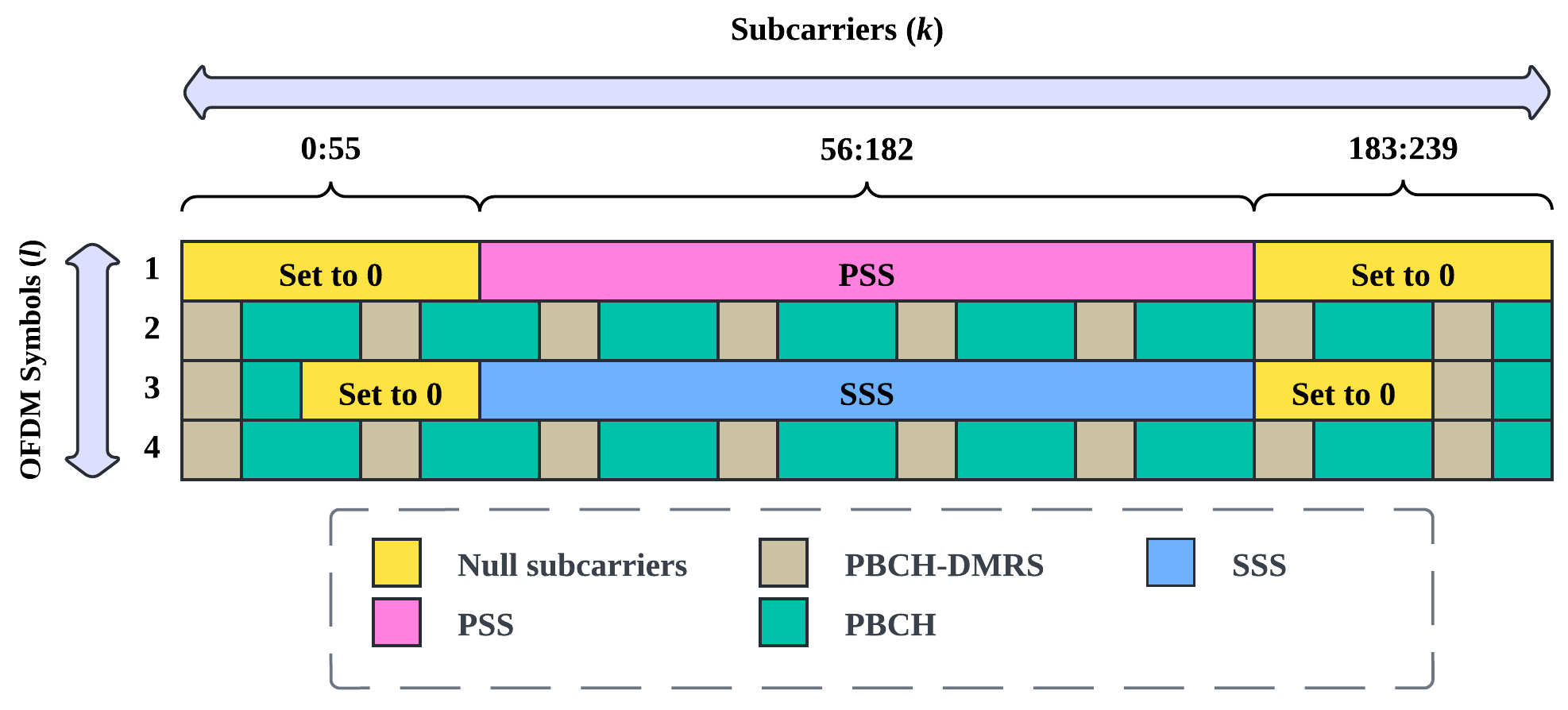}
        \caption{} 
        \label{fig:ssb}
    \end{subfigure}
    %\hfill
    \begin{subfigure}{0.29\textwidth}
        \centering
        \includegraphics[width=0.9\textwidth, height=4cm]{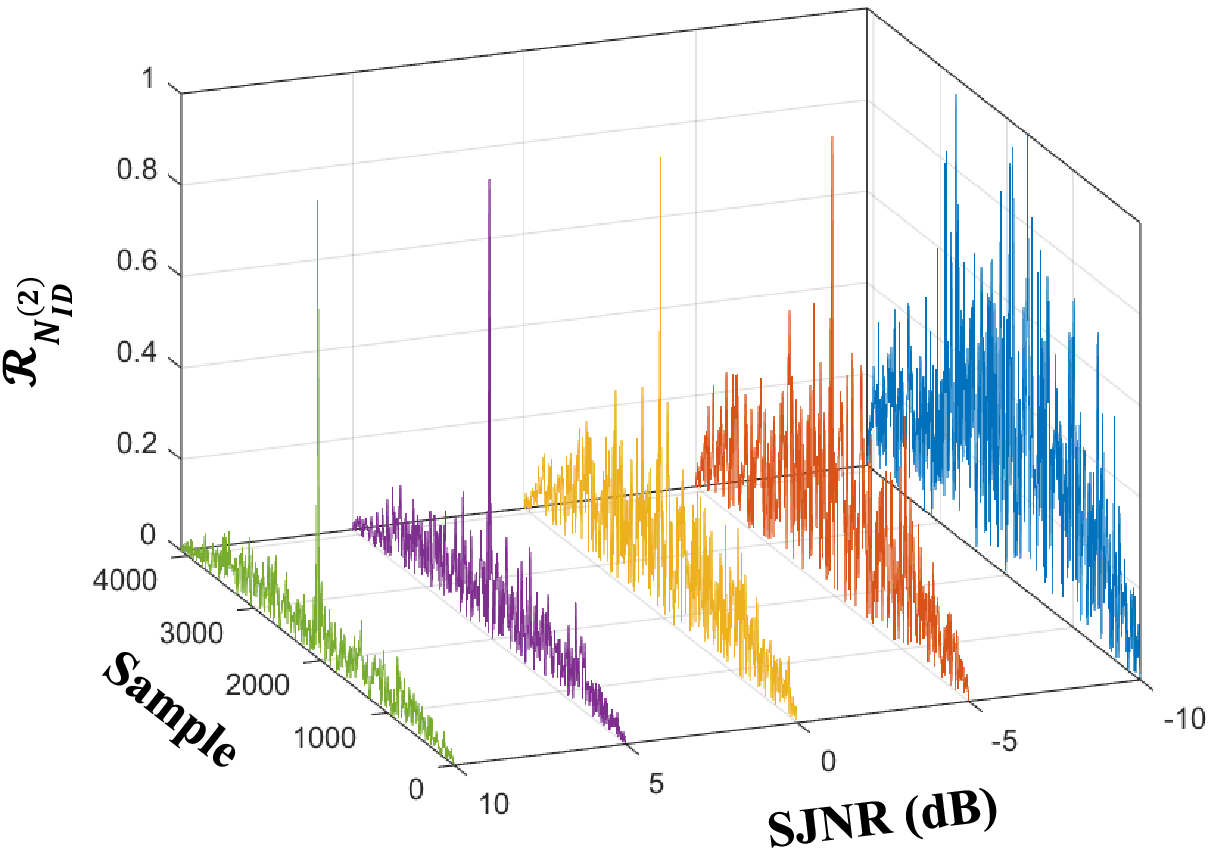}
        \caption{}
       \label{fig:psscorr}
    \end{subfigure}
        \begin{subfigure}{0.28\textwidth}
        \includegraphics[width=0.9\textwidth, height=4cm]{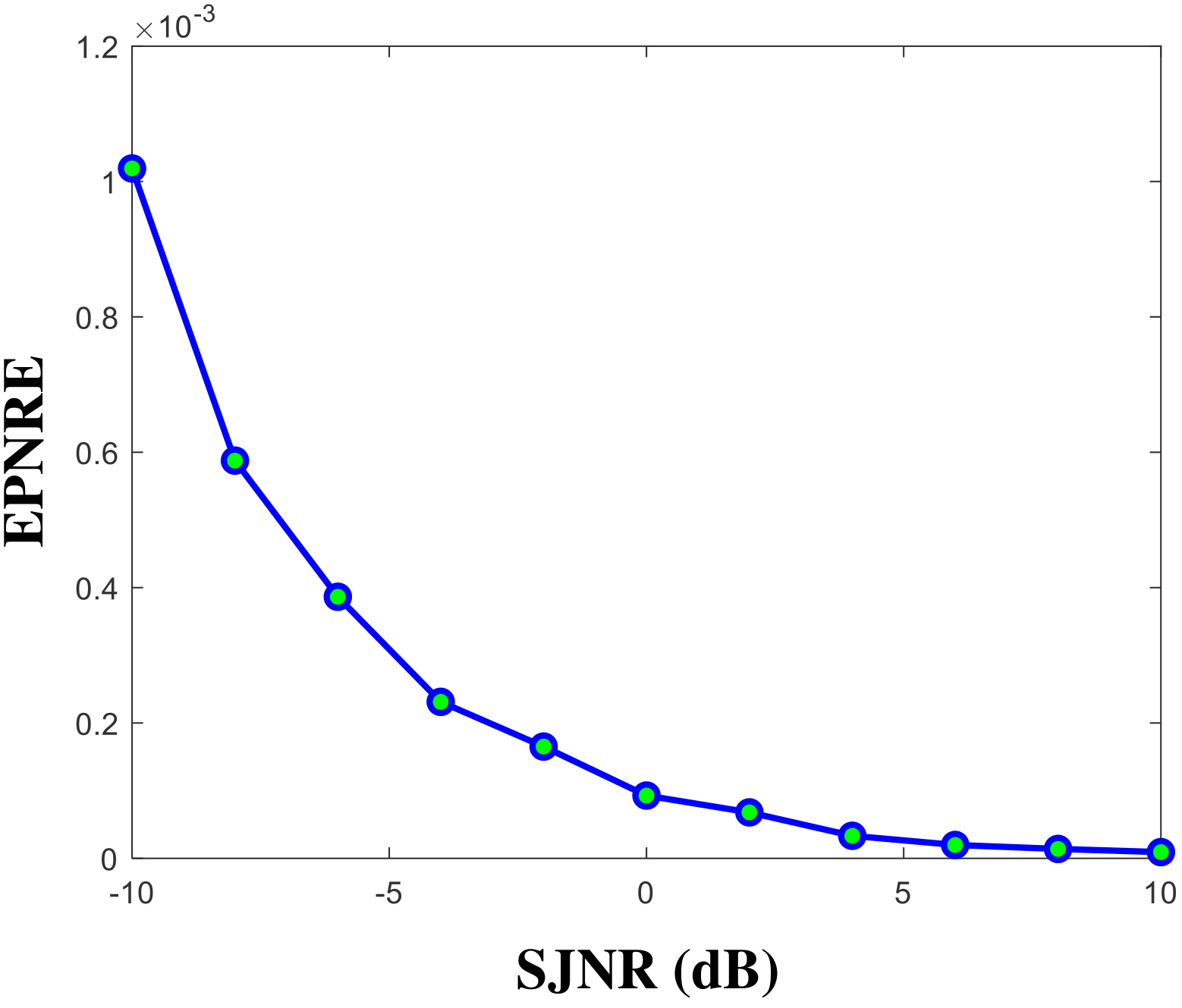}
        \caption{} 
        \label{fig:epnre}
    \end{subfigure}
    \hfill
   %\begin{subfigure}{0.23\textwidth}
        %\centering
        %\includegraphics[width=1\textwidth, height=4cm]{figures/fig_spec_barrage.pdf}
        %\caption{}
       %\label{fig.fig_PBCH_cons_boost}
    %\end{subfigure}

   \caption{(a) \small 5G Signal Synchronization Block (SSB). (b) \small The PCI detection process with cross correlation between the received PSS by UE and known basis correlation under different SJNR values - UE is $250\, m$ away from the gNB. (c) Energy per null resource elements under different SJNR values -UE is $50\, m$ away from the gNB.}
    \label{fig: feasibility} \vspace{-4mm}
\end{figure*}

The SSB, illustrated in Fig. \ref{fig:ssb}, is usually transmitted within the 5G radio frame using 4 OFDM symbols \cite{kopacz2021effective} and includes two synchronization signals: the PSS and SSS \cite{lin2018ss}. The SSB comprises information necessary for the UE to establish synchronization with the cell, such as the PCI of the cell, and additional information provided by the PBCH and physical downlink shared channel (PDSCH)\cite{lin2018ss}. 
PSS and SSS use the same time slots as the PBCH. PBCH symbols are concentrated in two or four slots, which gives the appearance of a low-duty cycle, particularly at higher subcarrier spacing. The MIB, containing critical data transported by the PBCH, comprises parameters that are vital for the UE to establish a connection with a cell.
When a UE powers on or enters a new cell's coverage area, it uses this information to discover the available SSBs. Scanning the predefined SSB locations within the radio frame, the UE executes an SSB discovery procedure to accomplish this. The UE is capable of decoding the information carried by an SSB, including the PCI, once it has been detected. 
This detection process is performed by analyzing the correlation of the received SSS and PSS signals and the known base SSS and PSS signals. A jammer that impacts the SSB of 5G communication can disrupt the PCI detection and synchronization process by a simple interference signal. PSS signals will be described and mathematically modelled in section \ref{section:psscorr}. However, to better understand the effect of the jamming attack on the synchronization process, we provide the output of PCI detection based on PSS, i.e. the cross correlation between received PSS and basis PSS sequences $\left(\mathcal{R}_{N_{ID}^{(2)}}(t)\right)$, under different jamming powers in Fig. \ref{fig:psscorr}. When the SJNR value is equal to $10\, dB$, the output of the correlation of received PSS and base PSS sequence results in a clear peak which leads the UE to detect the PCI of the available gNB. As the SJNR decreases and reaches $-10\, dB$ with higher jamming power, the correlation peak detection becomes more challenging, disrupting legitimate communication.
In addition, based on the SSB in Fig. \ref{fig:ssb}, it can be concluded that some resource elements are set to zero based on the 3GPP standard containing zero energy. When a jammer starts sending jamming signals, the energy of these resource elements starts increasing.
This is another characteristic that can help to detect jamming attacks as detailed under the Energy per Null RE (EPNRE) description in section \ref{section:epnre}. Fig. \ref{fig:epnre} demonstrates the total energy of all null resource elements when the power of jammer decreases and SJNR is increased from $-10\, dB$ to $10\, dB$. It is observed that as the jamming power is decreased, the EPNRE decreases toward zero. This change in EPNRE can be used to detect the presence of a jammer in the network. \vspace{-2.75mm} %Additionally, Fig. \ref{fig:spec-ssb} demonstrates the effect of barrage and smart SSB jamming on the RF domain 5G resource grid. The RF domain samples have been collected from the TELUS operator in the band n71. Fig. \ref{fig:spec-nojam} shows its spectrogram when there is no jamming signal. As annotated, two SSBs are found in that time interval. \ref{fig:spec-ssb} (b) and (c) shows the same spectrogram under barrage (SJNR=-5dB) and smart SSB (SJNR=0dB) attack, respectively. In case of the barrage jammer, the whole time-frequency map is affected by the the destructive jamming signal while smart SSB attempts to ... 

\graphicspath{{figs/}}
\begin{comment}
\graphicspath{{figs/}}
\begin{figure*}
    \centering
    \begin{subfigure}{0.30\textwidth}
        \includegraphics[width=1\textwidth, height=4cm]{figures/fig_Spec_no_jam.pdf}
        \caption{} 
        \label{fig:spec-nojam}
    \end{subfigure}
    %\hfill
    \begin{subfigure}{0.30\textwidth}
        \centering
        \includegraphics[width=1\textwidth, height=4cm]{figures/fig_spec_barrage.pdf}
        \caption{}
       \label{fig:spec-barrage}
    \end{subfigure}
        \begin{subfigure}{0.30\textwidth}
        \includegraphics[width=1\textwidth, height=4cm]{figures/fig_spec_SSB.pdf}
        \caption{} 
        \label{fig:spec-ssb}
    \end{subfigure}
    \hfill
   %\begin{subfigure}{0.23\textwidth}
        %\centering
        %\includegraphics[width=1\textwidth, height=4cm]{figures/fig_spec_barrage.pdf}
        %\caption{}
       %\label{fig.fig_PBCH_cons_boost}
    %\end{subfigure}

   \caption{\small 5G resource grid obtained from TELUS Network- NR n71 band (a) No jamming attack (b) Barrage Jamming attack SJNR=-5dB  (c) Smart SSB attack SJNR=0dB}
    \label{fig: spectogram}
\end{figure*}

\end{comment}
%------------------------System model Intro-------------------------------------
\subsection{Problem formulation}
This section provides an introduction to SSB, as the focus of our detection algorithm is on analyzing information derived from this block. The rest of this section presents a detailed explanation of the 5G system model, along with the formulation of the jamming detection problem. The notations used in this paper are defined in Table \ref{tab:symbols}.

\begin{table}[htbp]
\caption{Description of Notaions } \vspace{-3mm}
\label{tab:symbols}
\begin{center}
\begin{tabular}{ |c|c|c| }
\hline
Symbol & Description & Equation \\
\hline \hline
$\Phi_{N_{ID}}$ & PSS base sequences & \ref{eq: m-sequences}, \ref{eq: tx symbol 0} \\ \hline
$X_{l,k}^{ssb}$ & Transmitted SSB in frequency domain & \ref{eq: tx symbol 0}, \ref{eq: ifft} \\ \hline
$x_l^{ssb}(t)$ & Transmitted SSB in time domain & \ref{eq: ifft}, \ref{eq:rx ssb}, \ref{eq:pss corr}, \ref{eq:hypothesis}, \ref{eq:hypothesis2} \\ \hline
$y_l^{ssb}(t)$ & Received SSB in time domain& \ref{eq:rx ssb}, \ref{eq: rx pss}, \ref{eq:hypothesis}, \ref{eq:hypothesis2}, \ref{eq: fourier of rx}\\ \hline
\multirow{2}{4em}{$y_{pss}(t)$} & Extracted PSS sequence from the received & \ref{eq: rx pss}, \ref{eq:pss corr}\\ 
& signal in time domain & \\ 
\hline
$Y_{l,k}^{ssb}$ & Received SSB in frequency domain& \ref{eq: fourier of rx}, \ref{eq:energy}\\ \hline
\multirow{3}{4em}{$\boldsymbol{\mathcal{R}_{N_{ID}^{(2)}}}(t)$} & Correlation of extracted PSS sequence & \ref{eq:pss corr}, \ref{eq:DWT}\\ 
&from the received signal & \\
& and three m-sequences &\\
\hline
$\boldsymbol{\mathcal{R}^i_{N_{ID}^{(2)}}}$ & Output of $i^{th}$ DWT stage& \ref{eq:DWT}, \ref{eq:DWT2}, \ref{eq:final dataset}\\ \hline
\multirow{2}{4em}{$E, \boldsymbol{\mathcal{E}}$} & Energy per null resource elements, & \ref{eq:energy}, \ref{eq:final dataset}\\ 
& vectorization of log transform of energy & \\
\hline
\multirow{2}{4em}{\ \ \ \ $\boldsymbol{\chi}^i$} & 3-D tensor of $i^{th}$ observation & \ref{eq:final dataset}, \ref{eq: normalized z}, \ref{eq: decision},\\ 
& & \ref{eq: ML}, \ref{eq: negative LL}\\
\hline
$\boldsymbol{\mathcal{Z}}^i$ & Corresponding label for $i^{th}$ observation & \ref{eq: label set}, \ref{eq: ML}, \ref{eq: negative LL}\\ \hline
\multirow{2}{4em}{\ \ \ $\zeta_{H_{i}}$} & Score of the jamming detection algorithm& \ref{eq: normalized z}, \ref{eq: decision}, \ref{eq: ML},\\ 
&  for $i^{th}$ class &\ref{eq: negative LL}, \ref{eq: eta1}, \ref{eq: eta2}\\
\hline
$\gamma$ & DNN model threshold & \ref{eq: gamma2}, \ref{eq: final decision}\\
\hline
\end{tabular}  \vspace{-4mm}
\end{center}
\end{table}

In the first step of gaining access to a gNB, UE requires information called System Information Block 1 (SIB1) which depends on the decoding of MIB \cite{chen2020design}. This procedure is achievable only through the detection of SSB. One or more SSBs are transmitted through an SS burst periodically based on a pre-determined periodicity at a five-millisecond window \cite{3gpp.38.213}. Each SSB contains cell ID information which is calculated by $N_{ID}^{Cell}=3\times N_{ID}^{(1)}+N_{ID}^{(2)}$. Where $N_{ID}^{(1)}\in\{0,1,...,335\}$ represents group ID and $N_{ID}^{(2)}\in\{0,1,2\}$ is related to sector ID \cite{3gpp.38.211}. Sector ID is provided by PSS to help reach a coarse time and frequency synchronization. A 5G frame includes numerous slots, each being divided into a specific number of symbols which depends on the subcarrier spacing and numerology. SSB is transmitted through four OFDM symbols in the time domain and spans over $k_{ssb}=240$ subcarriers in the frequency domain. 127 subcarriers in the first symbol are dedicated to PSS, and there are 113 unused subcarriers below and above PSS which are set to '$0$'. PSS follows on of the three base $k_{pss}$-symbols m-sequences $\Phi_{N_{ID}^{(2)}}(k)$ in frequency domain. Each m-sequence is a circular shift version of the other two and their cross-correlation value is equal to zero \cite{you2020efficient}. The three base m-sequences are demonstrated in (\ref{eq: m-sequences}).\vspace{-7mm}

\begin{equation} \label{eq: m-sequences}
    \begin{split}
        \Phi_{N_{ID}^{(2)},k}&=1-2s(m)\\
        m=(k+43\times N^{(2)}_{ID})&mod \, k_{pss},0\leq k<k_{pss}
    \end{split}
\end{equation}

\noindent where $\Phi_{N_{ID}^{(2)},k}$ demonstrates the PSS symbol at subcarrier $k$, and $k_{pss}$ is the length of the PSS sequence in the frequency domain. Furthermore, $s(m)$ represents the m-sequences which are given as,\vspace{-2mm}
\begin{equation}\label{eq: sequence} 
\begin{split}
    s(i+7)=(s(i+4)+s(i)&) \ mod \, 2\\
        [s(6) \ s(5) \ s(4) \ s(3) \ s(2) \ s(1) \ s(0)] &= [1 \   1 \ 1 \ 0 \ 1 \ 1 \ 0]
\end{split}
\end{equation} 

% 5G Network Model-------------------------------------------------
Consider a gNB generating a waveform $X^{ssb}_{l,k}$ in frequency domain containing the SSB, where $l\in\{0,1,2,3\}$ denotes the OFDM symbol in 5G resource grid and $k\in\{0,1,...,k_{ssb}-1\}$ represents the subcarrier data point. The transmitted SSB includes the PSS sequence in the first OFDM symbol as defined in (\ref{eq: tx symbol 0}). \vspace{-6mm}

\begin{equation}\label{eq: tx symbol 0}
    X_{l,k}^{ssb}\big |_{l=0}=\begin{cases}
        \Phi_{N_{ID}^{(2)},k},\ k\in \{56, 57,...,182\}\\
        0,\ \ \ \ \ \ \ \ \ \ \ otherwise
    \end{cases}
\end{equation}
The signal is subjected to modulation and after applying IFFT, the time domain signal ($x^{ssb}_l(t)$\footnote{t here indicates the time index for a discrete-time signal.}) is formed as, \vspace{-3mm} 

\begin{equation}\label{eq: ifft}
\begin{split}
    x^{ssb}_l(t)=\frac{1}{k_{ssb}} \sum_{k=0}^{k_{ssb}-1}X_{l,k}^{ssb}e^{j\frac{2\pi}{k_{ssb}}tk} \quad
    l&\in\{0,1,2,3\}
\end{split}
\end{equation}

During the transmission, the transmitted signal experiences the impact of channel model and thermal noise, and it is degraded by the path loss attenuation. Therefore, the received SSB ($y_{l,k}^{ssb}(t)$) having been subjected to the impact of thermal noise ($\sigma_{th}$) and channel model ($h(t)$) is represented in (\ref{eq:rx ssb}) where $N_{FFT}$ represents the number of FFT points.\vspace{-3mm}

\begin{equation}\label{eq:rx ssb}
    y^{ssb}_l(t) = \sum_{\tau=0}^{N_{FFT}-1} x^{ssb}_l(\tau) h(t-\tau) + \sigma_{th}
\end{equation}

As the received PSS sequence is transmitted through the first OFDM symbol ($l=0$), it can be extracted by taking the first $N_{FFT}$ samples of the time domain signal as, \vspace{-3mm}

\begin{equation}\label{eq: rx pss}
\begin{split}
    y_{pss}(t) = y_{l}^{ssb}(t)\big|_{l=0}, \quad
    t\in\{&0,...,N_{FFT}-1\}
\end{split}
\end{equation}

In the scenario where a jamming signal ($x_j(t)$) \footnote{As the focus of our detection system is investigating the impact of the jammer on 5G SSB, the jamming signal is modeled based on the AWGN noise that reflects both smart SSB jammer and barrage jammer.} is introduced, the received signal becomes susceptible to the impact of the jammer. Detection of the jamming signal can be represented as a binary hypothesis framework (\ref{eq:hypothesis}),(\ref{eq:hypothesis2}), where we have a null hypothesis denoted as $H_0$ and an alternative hypothesis, $H_1$, corresponding to the presence of the jamming signal.\vspace{-3mm}

\begin{numcases}{}\label{eq:hypothesis}
    H_0 : & $y^{ssb}_l(t) = \sum_{\tau} x^{ssb}(\tau) h(t-\tau) + \sigma_{th} $ \\ \label{eq:hypothesis2}
    H_1 : & $y^{ssb}_l(t) = \sum_{\tau} x^{ssb}(\tau) h(t-\tau) + \sigma_{th} + x_j[t]$
\end{numcases}
% path loss: d^-v

Let $\hat{H}$ be the output of the test based on the observations. Two main performance metrics used in this work are the probability of jamming detection ($P_D=P_r(\hat{H}=H_1|H_1)$), and the probability of false alarm ($P_{FA}=P_r(\hat{H}=H_1|H_0)$)

\section{Proposed DNN-based jammer detection}
\label{detector}
\begin{figure*}
    \centering
    \includegraphics[width=.9\textwidth, height=8cm]{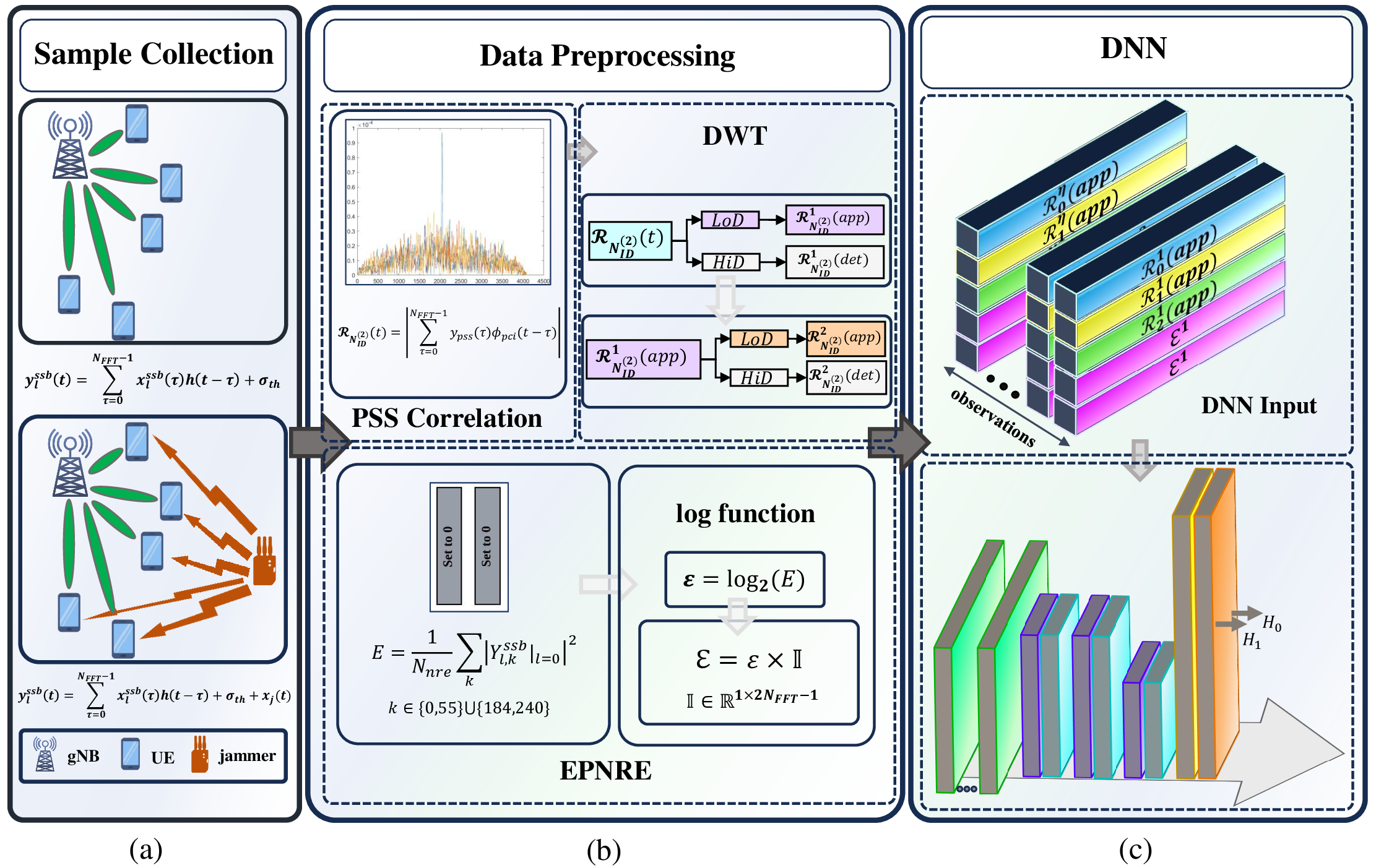}
    \caption{\small The block diagram of the proposed detection technique using information from SSB}
    \label{fig:block diagram} \vspace{-4 mm}
\end{figure*}

This section focuses on the proposed jamming detection scheme. The overall block diagram of our detection model is demonstrated in Fig. \ref{fig:block diagram}.
Received signal observations are processed through a data preprocessing module, following which the modified dataset is input into a Deep Learning block for training and jamming detection. The details of each block are provided in the rest of this section.

\graphicspath{{figures/}}
% ___________Preprocessing___________
\subsection{Data Preprocessing}
Preprocessing of data in ML techniques is a vital part. If done appropriately, it helps transfer raw data into a neat feature space that can be exploited by the DL module for high-performance classification. Data preprocessing in the proposed architecture includes three main steps, performing PSS correlation, calculating DWT, and extracting EPNRE. The first step helps detect low-power jamming since the peak amplitude and the correlation output pattern can reflect the presence of the jammer even in small jamming powers as depicted in Fig. \ref{fig:psscorr}. The DWT acts as a feature space dimension reduction techniques, making the training process more easy and less time-consuming. Extracting EPNRE helps the DL decision be robust to channel behavior as no signal energy from the gNB is assumed to be there.
\subsubsection{PSS correlation}
\label{section:psscorr}
Each transmitted SSB includes a PSS sequence which is a version of one of the three base PSS sequences formulated in (\ref{eq: m-sequences}). Thus, within the correlation signal of the received PSS sequence $y_{pss}(t)$ with one of the m-sequences, we expect to observe a visible peak value.
The correlation of $y_{pss}(t)$ with each m-sequences results in a 1 by 2$N_{FFT}$ signal demonstrated as\footnote{Since there is a specific $x^{ssb}_l(t)$ signal for each $N_{ID}^{(2)}$, the output of the correlator in (\ref{eq:pss corr}) is sub-scripted with $N_{ID}^{(2)}$. }, \vspace{-7mm}

\begin{equation}\label{eq:pss corr}
\begin{split}
    \boldsymbol{\mathcal{R}_{N_{ID}^{(2)}}}(t)& = \Bigg|\sum_{\tau=0}^{N_{FFT}-1}y_{pss}(\tau)x^{ssb}_l(t-\tau) \big|_{l=0} \Bigg|,\\
\end{split}
\end{equation} where $|.|$ denotes the absolute value. By analyzing the correlation sequences, the cell ID parameter of the correspondent gNB can be specified. The jammer can distort this cell ID extraction process through jamming signals. 

\subsubsection{DWT}

\begin{figure}[htp]
    \centering
    \includegraphics[width=.7\linewidth, height=5cm]{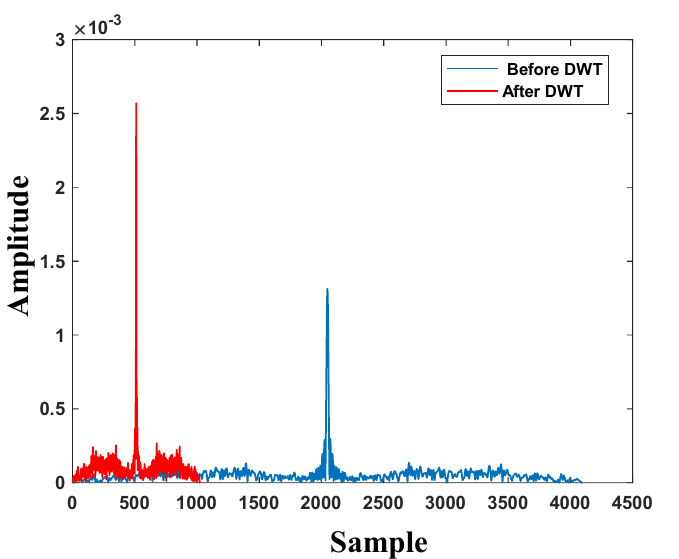}
    \caption{\small Applying DWT to the original dataset. The blue line is the signal before DWT and red line demonstrates the effect of DWT on the signal.}
    \label{fig:dwt effect}
\end{figure}

To remove the redundant information from the correlator output and to magnify the jammer-related features, DWT is employed. This also reduces the dimension of the data set and hence, significantly decreases the training time as described in \ref{Sec: Single Threshold}. As illustrated in Fig. \ref{fig:block diagram}, two-stage DWT is applied to the output of the correlator, $\boldsymbol{\mathcal{R}_{N_{ID}^{(2)}}}(t)$. 
\vspace{1mm}
The approximate output of the first DWT module, $\boldsymbol{\mathcal{R}^1_{N_{ID}^{(2)}}}(app)$, 
\vspace{1mm}
is fed to the second DWT module, generating the final outputs $\Big(\boldsymbol{\mathcal{R}^2_{N_{ID}^{(2)}}}(app) , \, \boldsymbol{\mathcal{R}^2_{N_{ID}^{(2)}}}(det)\Big)$. 
\vspace{1mm}
This process can be mathematically explained as follows: \vspace{-4mm}

\begin{equation}\label{eq:DWT2} 
\begin{split}
    \big(\boldsymbol{\mathcal{R}^2_{N_{ID}^{(2)}}}(app) , \, \boldsymbol{\mathcal{R}^2_{N_{ID}^{(2)}}}(det)\big)=\\
    & \hspace{-15mm}  \frac{1}{\sqrt{2N_{FFT}}} \sum_m \boldsymbol{\mathcal{R}^1_{N_{ID}^{(2)}}}(app)\psi_{app,det}(t) ,
\end{split}
\end{equation}
\noindent in which $\boldsymbol{\mathcal{R}^1_{N_{ID}^{(2)}}}(app)$ is obtained as, \vspace{-3mm}

\begin{equation}\label{eq:DWT}
\begin{split}
   \big(\boldsymbol{\mathcal{R}^1_{N_{ID}^{(2)}}}(app) , \, \boldsymbol{\mathcal{R}^1_{N_{ID}^{(2)}}}(det)\big)= \\
   & \hspace{-15mm} \frac{1}{\sqrt{2N_{FFT}}}\sum_m \boldsymbol{\mathcal{R}_{N_{ID}^{(2)}}}(t)\psi_{app,det}(t).
\end{split}  
\end{equation}
where $\boldsymbol{\mathcal{R}_{N_{ID}^{(2)}}}(t)$ is derived as per (\ref{eq:pss corr}).
 Furthermore, $\psi_{app,det}(t)$ is the mother wavelet function chosen as Haar wavelet \footnote{Haar wavelet is a set of rescaled square-shaped functions that collectively form the simplest wavelet family or basis \cite{stankovic2003haar}.} to maintain and magnify the important information from PSS correlation. Haar wavelet is a straightforward choice as it is very effective at identifying abrupt changes in the signal amplitude including PSS peaks \cite{liu2021applications}\cite{tang2018haar}. Thus, it can successfully reduce signal size while maintaining crucial information and keeping the computational load at the desired stage. Fig. \ref{fig:dwt effect} demonstrates the effect of adding two layers of DWT to the PSS correlation dataset. As can be seen in the figure, the length of the correlated signal is decreased by almost 25\%, and the correspondent peak is magnified significantly by 2 times. 
\begin{equation}
    \psi(t)=\begin{cases}
    1,\ \ \ &0\leq t <\frac{1}{2},\\ 
    -1,\ \ \ &\frac{1}{2}\leq t<1\\
    0,\ \ \ &otherwise
\end{cases}
\end{equation}
\vspace{-6mm}
\graphicspath{{figures/}}

\subsubsection{EPNRE}
\label{section:epnre}
After receiving the transmitted signal, the signal is converted into the frequency domain using Fourier transform as in (\ref{eq: fourier of rx}). \vspace{-2mm}
\begin{equation}\label{eq: fourier of rx}
    Y_{l,k}^{ssb}=\frac{1}{N_{FFT}}\sum_{t=0}^{N_{FFT}-1}y_l^{ssb}(t)e^{-j\frac{2\pi}{N_{FFT}}k t}
\end{equation}
\noindent in which $Y_{l,k}^{ssb}$ represents the Fourier transform of the received signal in $l$-th OFDM symbol and $k$-th subcarrier data point.
In the next step, the SS block is extracted from the frequency domain signal. There are $N_{nre}=113$ numbers of resource elements in the SS block which are intentionally set to zero by gNB \cite{you2020efficient}. These resource elements help us collect information about the noise and jamming signal. The energy of these resource elements is calculated as expressed in (\ref{eq:energy}). \vspace{-4mm}

\begin{equation}\label{eq:energy}
\begin{split}
    E =  \frac{1}{N_{nre}}\sum_{k} \Big|{Y_{l,k}^{ssb}\mid_{l=0}} \Big|^2 \quad
    k&\in{\{0,55\}}\cup{\{184,240\}}
\end{split}
\end{equation}
Depending on the jammer transmit power and its distance to the UE, the received energy of the jamming signal might be small in some observations. Thus, to adjust the dynamic range of the calculated received energy to be more distinguishable by the classifier, a log transform is applied to the final energy value, i.e., $\varepsilon = \log_2(E)$.

In the absence of a jammer, the UE is only affected by environmental noise. The energy of this type of interference is significantly small compared to the energy from a jammer. Thus, when the $\varepsilon$ reaches a specific threshold, it can be concluded that a jammer is present in the scenario.
\vspace{-5mm}
% ___________DNN Model___________ 
% DNN model structure
%% optimization problem
%% hyper parameter set table (minibatch,...)
\subsection{DL Block}
\subsubsection{DNN Input}
Observations are designed as a 2-D image including the PSS correlation signals at the output of the second DWT module ($\boldsymbol{\mathcal{R}^2_{N_{ID}^{(2)}}}(app), \, N_{ID}^{(2)}\in\{0,1,2\}$), and energy per null resource elements ($\varepsilon$).
In each observation ($N_{obs}\in\{1,2,...,2\mu_j\}$), the first three rows in the 2-D image include correlated signal, and the last two rows are dedicated to the energy parameter (Fig. \ref{fig:block diagram}). The calculated EPNRE is a scalar value while the size of $\boldsymbol{\mathcal{R}^2_{N_{ID}^{(2)}}}(app)$ is equal to $N_{FFT}/2$. Thus, to maintain the impact of the energy feature and to make it visible to the classifier, the energy is repeated in the last two rows. Mathematically speaking, $ \boldsymbol{\mathcal{E}} = \varepsilon\times{\mathbb{I}}$ where $\mathbb{I}$ denotes a $(1 \times N_{FFT}/2)$ vector of ones. 
After creating a 2-D image of features, each observation is positioned behind the previous observation thus creating a 3-D tensor (Fig. \ref{fig:block diagram}) as the final dataset ($\boldsymbol{\chi}$).\vspace{-4mm}

\begin{equation}\label{eq:final dataset}
    \boldsymbol{\chi}=[\boldsymbol{\mathcal{R}^2_0}^T(app),\boldsymbol{\mathcal{R}^2_1}^T(app),\boldsymbol{\mathcal{R}^2_2}^T(app),\boldsymbol{\mathcal{E}}^T,\boldsymbol{\mathcal{E}}^T]^T
\end{equation}

\subsubsection{Data Augmentation and Class Balancing}
%During the data generation, a noticeable dissimilarity in sample sizes between the two classes has been observed 
The dissimilarity in sample sizes between the two classes causes the class imbalance that can result in biased models and fail to generalize to minority class \cite{buda2018systematic}. Data augmentation techniques provided in several works, notably in \cite{mumuni2022data, shorten2019survey, maharana2022review} have demonstrated efficacy in enhancing the precision of classification tasks involving unbalanced datasets \cite{cai2022spectrum}. Circular shift is one of the data augmentation methods applied to the minor dataset to overcome the effect of the imbalance classes and is introduced and described in detail in \cite{zhang2020circular}. Another reason for performing circular shift is that the PSS correlation peak in (\ref{eq:pss corr}) might not happen exactly at the center sample of $\boldsymbol{\mathcal{R}_{N_{ID}^{(2)}}}(t)$ due to imperfect synchronization in time domain and sample rate mismatch between gNB and the receiver. Circular shift helps the DNN model to learn different patterns existed in the real world data.
Through the circular shift augmentation method, each observation in the minority class is divided into sub-sequences with similar lengths, and each sub-sequence is independently shuffled to create a new sequence which introduces new variations of the collected observations.
Traditional techniques (such as rotation, cropping, and flipping) may affect data integrity or eliminate vital information. Circular shift enhances the generalization capabilities of the DNN model by introducing diverse patterns while maintaining most of the original information without significant information loss  \cite{zhang2020circular}, \cite{singh2023systematic}.

\subsubsection{DNN Structure}
The deep learning model used for training is a supervised model, and dataset is paired with the correspondent labels as $(\boldsymbol{\chi}, \boldsymbol{\mathcal{Z}})=\{(\boldsymbol{\chi}^1,\boldsymbol{\mathcal{Z}}^1),(\boldsymbol{\chi}^2,\boldsymbol{\mathcal{Z}}^2),...,(\boldsymbol{\chi}^{\eta},\boldsymbol{\mathcal{Z}}^{\eta})\}$, where $\eta$ is the total number of observations. In this algorithm two classes are represented for the dataset, therefore, $\boldsymbol{\mathcal{Z}}^{\eta}$ represents a binary digit set and can be written using one-hot encoding (\ref{eq: label set}). \vspace{-3mm}

\begin{equation}\label{eq: label set}
    \boldsymbol{\mathcal{Z}}^{\eta}=\begin{cases}
        [0\ \ 1]^T\ \ \ \ for\ H_0\\
        [1\ \ 0]^T\ \ \ \ for\ H_1
    \end{cases}
\end{equation}
After normalizing using the softmax layer, the output score of the model for $\eta$-th observation is expressed as follows:
\begin{equation}\label{eq: normalized z}
    \boldsymbol{O}^{\eta}=\begin{cases}
        \zeta_{H_0}(\chi^\eta|\theta)\ \ \ \ for\ H_0\\
        \zeta_{H_1}(\chi^\eta|\theta)\ \ \ \ for\ H_1
    \end{cases}
\end{equation}
\noindent where $\theta$ represents the parameter of the deep learning model, and $\zeta_{H_{i}}(\chi^\eta|\theta)\triangleq P_r\big(\chi^\eta|H_{i},\theta\big), i\in{0,1}$ denotes the jamming scores of the jamming detection problem which satisfies $\zeta_{H_{0}}(\chi^\eta|\theta)+\zeta_{H_{1}}(\chi^\eta|\theta)=1$. Based on the decision rule, two scores are compared and the hypothesis with the higher score becomes the output (\ref{eq: decision}). \vspace{-4mm}

\begin{equation}\label{eq: decision}
    \zeta_{H_{0}}(\chi^\eta|\theta)\underset{H_1}{\overset{H_0}{\gtrless}}\zeta_{H_{1\vee1}}(\chi^\eta|\theta)
\end{equation}

\begin{figure}[htp]
    \centering
    \includegraphics[width=8cm, height=4.5cm]{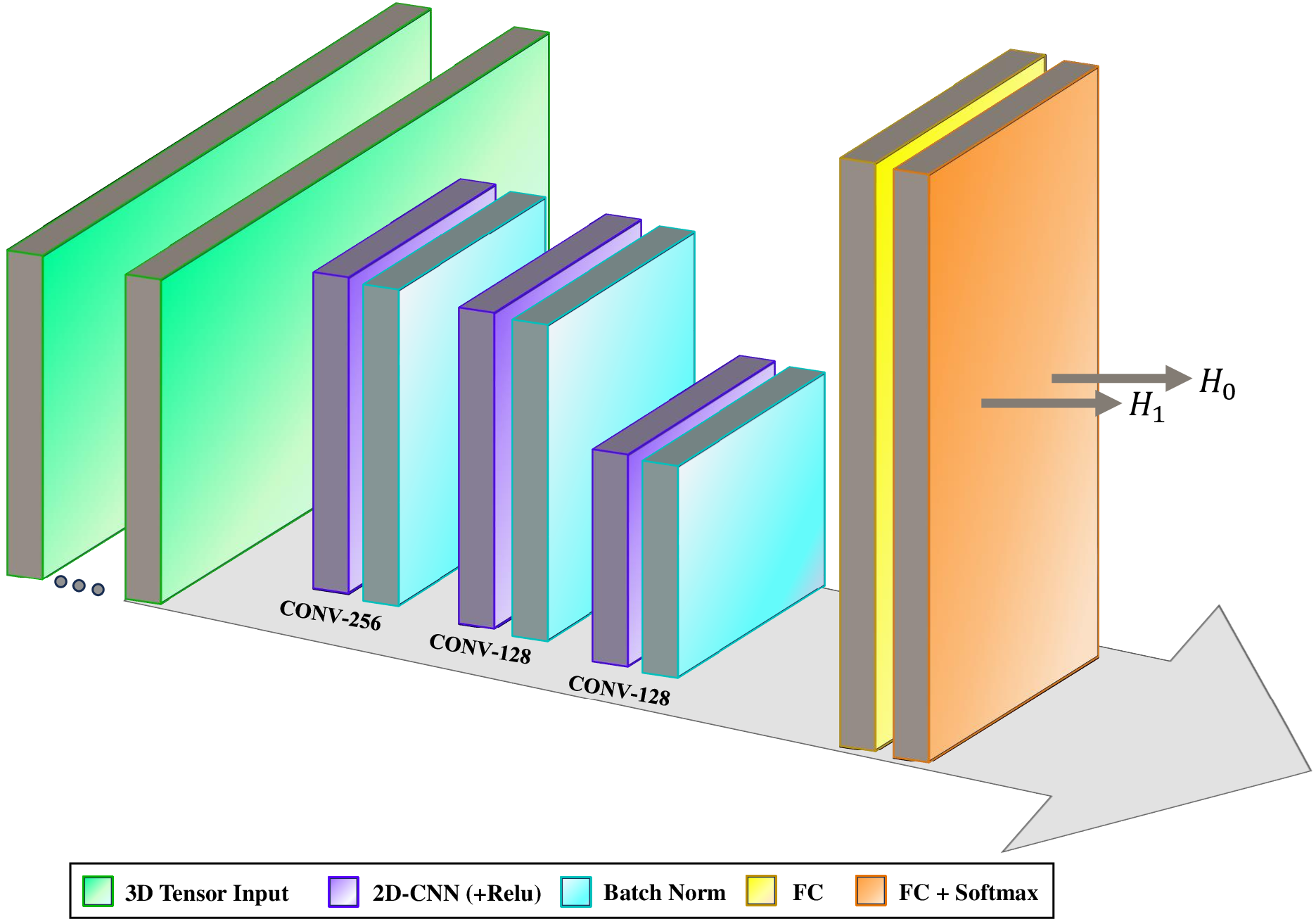}
    \caption{\small DNN model.} 
    \label{fig:dnn} \vspace{-5 mm}
    \vspace{2mm}
 \end{figure}

As the features are organized in 2-D images, the deep learning model is designed using 2-D CNN layers. After exploring various architectures and experimenting under different model parameters, the optimal design for the DNN is finalized as demonstrated in Fig. \ref{fig:dnn}. The 3-D tensors including the features of each observation are used as an input to three layers of 2-D CNN with ReLu activation. After each layer of CNN, a batch normalization layer is added to provide faster and more stable training. Two fully connected layers and a softmax layer are used as the last layers of the design. The softmax layer enables the normalization of the output scores. A detailed list of parameters for the DNN model is provided in Table \ref{tab:hyperparameter}. \vspace{-1 mm}

\begin{table}[htbp]
\caption{Hyperparameter Setting }
\vspace{-3mm}
\begin{center}
\begin{tabular}{|c|c|}
\hline
Parameter & Value \\
\hline\hline
Mini-Batch Size & 25\\
\hline
Initial Learning Rate & 0.001\\
\hline
MaxEpoch & 20\\
\hline
Validation frequency & 80\\
\hline
CNN-1 & $(2\times5) @ 256$\\
\hline
CNN-2 & $(2\times5) @ 128$\\
\hline
CNN-3 & $(1\times2) @ 128$\\
\hline
FC-1 Output Layer & 128\\
\hline
FC-2 Output Layer & 2\\
\hline
Training Optimization Method & SGDM\\
\hline
Validation Training Rate & 30\% \\
[1ex]
\hline
\end{tabular}  
\end{center}
\label{tab:hyperparameter}
\end{table}

\graphicspath{{figures/}}
%\vspace{-5mm}

\subsubsection{Offline Training}
% Math explanation
In the classification layer, the main goal is to maximize the probability of detection. Thus, maximizing $\zeta_{H_{0}}(\chi^\eta|\theta)$ in the absent of the jammer and $\zeta_{H_{1}}(\chi^\eta|\theta)$ in the presence of the jammer. To optimize the solution, an objective function can be defined as the Maximum Likelihood (ML) function represented in (\ref{eq: ML}). \vspace{-4mm}

\begin{equation}\label{eq: ML}
    L(\theta)= \prod_{\eta=1}^{N_{obs}} \big(\zeta_{H_{0}}(\chi^\eta|\theta)\big)^{\boldsymbol{\mathcal{Z}}^{\eta}}\big(\zeta_{H_{0}}(\chi^\eta|\theta)\big)^{1-\boldsymbol{\mathcal{Z}}^{\eta}}
\end{equation}

To minimize the loss between the actual and predicted value, a negative log-likelihood function is minimized to adjust the weights during the training as shown below:
\begin{equation}\label{eq: negative LL} \vspace{-3mm}
\begin{split}
    \mathcal{H}(\theta)&=-\frac{1}{N_{obs}}\log(l(\theta))=-\frac{1}{N_{obs}}\sum_{\eta=1}^{N_{obs}} \boldsymbol{\mathcal{Z}}^{\eta} \log\big(\zeta_{H_{0}}(\chi^\eta|\theta)\big)\\
    & \hspace{6 mm} + \>   (1-\boldsymbol{\mathcal{Z}}^{\eta})\log\big(\zeta_{H_{1}}(\chi^\eta|\theta)\big)
\end{split}
\end{equation}
Therefore, the optimization problem can be expressed as (\ref{eq: optimization dnn1}). 
\begin{equation}\label{eq: optimization dnn1}
    \Theta = \arg\min_{\theta}\mathcal{H}(\theta)
\end{equation}
Through stochastic gradient descent (SGD) optimization, the model is trained to provide higher values for $\zeta_{H_{0}}(\chi^\eta|\Theta)$ when $\boldsymbol{\mathcal{Z}}^{\eta}=[1\ \ 0]^T$ or $\zeta_{H_{1}}(\chi^\eta|\Theta)$ when $\boldsymbol{\mathcal{Z}}^{\eta}=[0\ \ 1]^T$. The training is formed through iterative back-propagating processes.
\vspace{-3mm}

% ___________Double Threshold___________
\subsection{Accuracy Enhancement}
% histogram
% double threshold model \gamma_1, \gamma_2 (D)
\subsubsection{Double Threshold DNN}
\begin{figure}[htp]
    \centering
    \includegraphics[width=8.5cm, height=2.5cm]{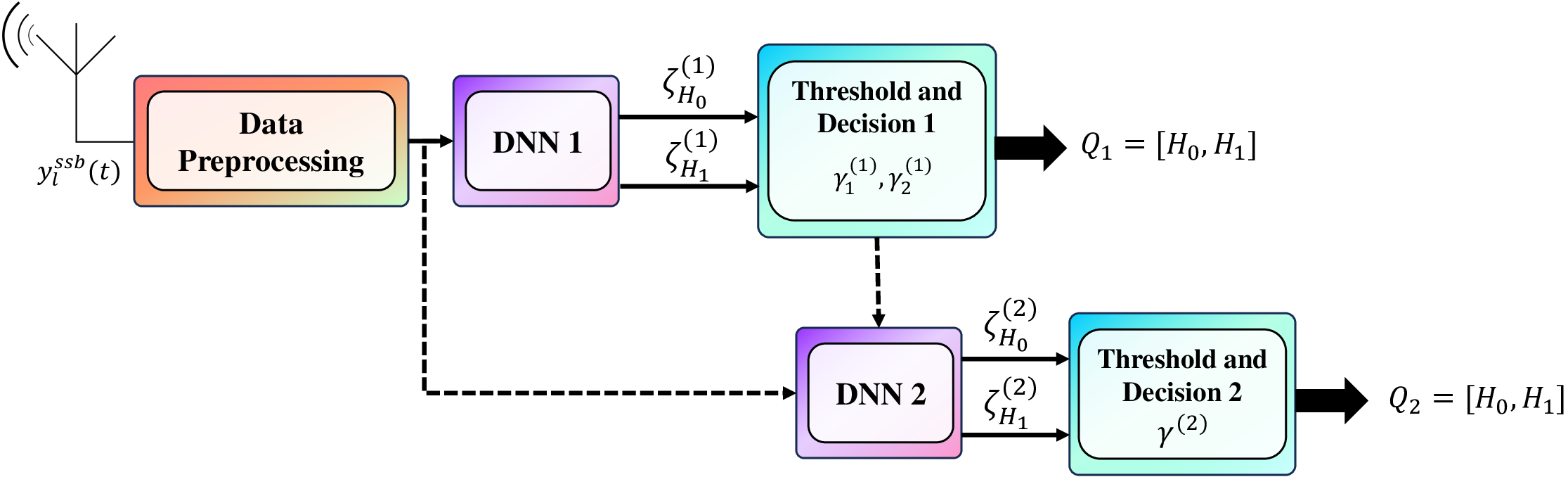}
    \caption{\small Adding another DNN block to improve the accuracy.} 
    \label{fig:dnn2}
    \vspace{-3mm}
 \end{figure}

The ratio between the two output scores of the detection model can be defined as $\Gamma(\chi^\eta)=\frac{\zeta_{H_1}(\chi^\eta|\Theta)}{\zeta_{H_0}(\chi^\eta|\Theta)}$. This value is compared with a threshold to enable the algorithm to make a detection decision. However, for the scenarios where SJNR value is high, meaning that the jamming signal is weaker compared to the transmitted signal from gNB, the uncertainty at the output of the classifier is high. Thus, the output scores are too close to each other, making it challenging for the classifier to make a decision. In other words, defining one hard decision at the output layer of the classifier reduces the accuracy of the jamming detector in the uncertainty area. To optimize the performance of the detection algorithm, this trade-off between the $P_D$ and $P_{FA}$ should be considered. Therefore, an ambiguity area is defined to reach a better and more accurate decision. Based on this, while the decision is being made out of the ambiguity area, the detector detects the absence or the presence of the jammer with 100\% empirical accuracy, balancing between the requirement for high detection rates against the risk of false alarms. To design such structure, two threshold points ($\gamma^{(1)}_1=\Gamma^{(1)}\big(\boldsymbol{\chi}_s^{\eta_1^*}\big), \gamma^{(1)}_2=\Gamma^{(1)}\big(\boldsymbol{\chi}_s^{\eta_2^*}\big)$) are considered. $\boldsymbol{\chi}_s$ represents sorted observations in descending order, and $\eta_1^*, \eta_2^*$ are observations that fall in the areas that the classification decides with high certainty. These thresholds are obtained as,
\begin{numcases}{}\label{eq: eta1}
        \eta^*_1=\arg\min_{\eta}\big(\zeta^{(1)}_{H_0}(\boldsymbol{\chi}_s^\eta|\Theta)<\zeta^{(1)}_{H_1}(\boldsymbol{\chi}_s^\eta|\Theta)\big)\\
        \label{eq: eta2}
        \eta ^*_2=\arg\max_{\eta}\big(\zeta^{(1)}_{H_0}(\boldsymbol{\chi}_s^\eta|\Theta)>\zeta^{(1)}_{H_1}(\boldsymbol{\chi}_s^\eta|\Theta)\big)
\end{numcases}
While $\Gamma^{(1)}(\chi^\eta)<\gamma^{(1)}_1$ or $\Gamma^{(1)}(\chi^\eta)>\gamma^{(1)}_2$, the detector selects $H_0$ or $H_1$ respectively. In the case that $\gamma^{(1)}_1<\Gamma^{(1)}(\chi^\eta)<\gamma^{(1)}_2$, another DNN is trained specifically for higher SJNR values is activated and observations are fed for more accurate analysis to this DNN. This improves the performance of the jamming detection by minimizing the probability of miss-detection (false negatives(FN)) and false-alarms (false positives(FP)), or equivalently, maximizing true positives (TP) and true negatives (TN). The structure of the DT-DDNN model is represented in Algorithm 1. After data is processed in the data preprocessing block, it is fed into the first DNN with a double threshold design. The first DNN calculates $\zeta^{(1)}_{H_0}(\boldsymbol{\chi}_s^\eta|\Theta)$ and $\zeta^{(1)}_{H_1}(\boldsymbol{\chi}_s^\eta|\Theta)$ which are the weight scores correspondence to this block. These scores are then used at the input of the first threshold and decision block (i.e., Threshold and Decision I). If the scores represent values outside of the uncertainty area, the final decision $Q_1$ is provided as an output. In the case that the score values fall into the uncertainty area, data is sent to the second DNN with a single threshold design for final classification. The second DNN is specifically trained in high SJNR regime to extract more fine-grained features and is expected to yield higher accuracy compared to the first DNN. In the second DNN, the threshold $\gamma^{(2)}$ is used to detect the jammer in harsher conditions (i.e., higher SJNR values). The ratio between the two output scores of the second DNN, $\zeta^{(2)}_{H_1}(\hat{\chi})$ and $\zeta^{(2)}_{H_0}(\hat{\chi})$, is defined as $\Gamma^{(2)}(\chi^\eta)=\frac{\zeta^{(2)}_{H_1}(\hat{\chi})}{\zeta^{(2)}_{H_0}(\hat{\chi})}$. The comparison between the ratio $\Gamma^{(2)}(\chi^\eta)$ and detection threshold $\gamma^{(2)}$ enables a degree of freedom to have a trade-off between $P_D$ and $P_{FA}$. After finding the $\Gamma^{(2)}(\chi^\eta)$ using optimum parameters of $\Theta$ and the $H_0$, empirical $P_{FA}$ can be computed to determine the threshold value. Assume we define the target value for $P_{FA}$ as $\delta_{FA}$ so that $P_{FA} \leq \delta_{FA}$, and we define the total number of observations fallen into $H_0$ hypothesis as $N_{H_0}$. After sorting the observations in descending order ($\chi_s^\eta$), the $\gamma^{(2)}$ threshold can be defined as in (\ref{eq: gamma2}). 
\begin{equation}\label{eq: gamma2}
    \gamma^{(2)}=\Gamma^{(2)}\big(\chi_s^\eta \lfloor \delta_{FA} N_{H_0} \rfloor;\Theta \big)
\end{equation}
The classification is processed based on comparing the ratio of the scores ($\Gamma^{(2)}(\chi^\eta)$) and single threshold value ($\gamma^{(2)}$) in the second DNN. During the online detection, samples are used as input to the DNN block, and after evaluations and passing through the softmax layer, two output scores are presented. The final decision is made in threshold and decision block using $\gamma^{(2)}$ as below: \vspace{-5 mm}

\begin{equation}\label{eq: final decision}
    \Gamma^{(2)}(\chi_{online})\underset{H_0}{\overset{H_1}{\gtrless}}\gamma^{(2)}
\end{equation}
\noindent in which $\chi_{online}$ represents the online samples. Fig. \ref{fig:dnn2} provides an overall overview of the structure of the double threshold enabled design explained above. %\vspace{-5 mm}

\graphicspath{{figures/}}
\vspace{-3mm}

\begin{algorithm}
\caption{DT-DDNN - Online Detection}\label{alg:cap}
\begin{algorithmic}[1]
\State Set $\hat{\chi}$ as the current observation
\State Set $\gamma^{(1)}_1=\Gamma\big(\hat{\chi}), \gamma^{(1)}_2=\Gamma\big(\hat{\chi})$
\State Calculate $\zeta^{(1)}_{H_0}(\hat{\chi}), \zeta^{(1)}_{H_1}(\hat{\chi})$
\State Calculate $\Gamma^{(1)}=\frac{\zeta^{(1)}_{H_1}(\hat{\chi})}{\zeta^{(1)}_{H_0}(\hat{\chi})}$

\If{$\Gamma^{(1)}<\gamma^{(1)}_1$}
    \State The classification decides $H_0$
\ElsIf{$\Gamma^{(1)}>\gamma^{(1)}_2$}
    \State The classification decides $H_1$
\ElsIf{$\gamma^{(1)}_1<\Gamma^{(1)}<\gamma^{(1)}_2$}
    \State Activate DNN-2
    \State Calculate $\zeta^{(2)}_{H_0}(\hat{\chi}), \zeta^{(2)}_{H_1}(\hat{\chi})$
    \State Calculate $\Gamma^{(2)}=\frac{\zeta^{(2)}_{H_1}(\hat{\chi})}{\zeta^{(2)}_{H_0}(\hat{\chi})}$
    \If{$\Gamma^{(2)}<\gamma^{(2)}$}
      \State The classification decides $H_0$
    \ElsIf{$\Gamma^{(2)}\geq \gamma^{(2)}$} 
      \State The classification decides $H_1$
    \EndIf  
\EndIf
%\EndWhile
\end{algorithmic} 
\end{algorithm}

 \subsubsection{Deep Cascade Learning}
\begin{figure}[htp]
    \centering
    \includegraphics[width=\linewidth]{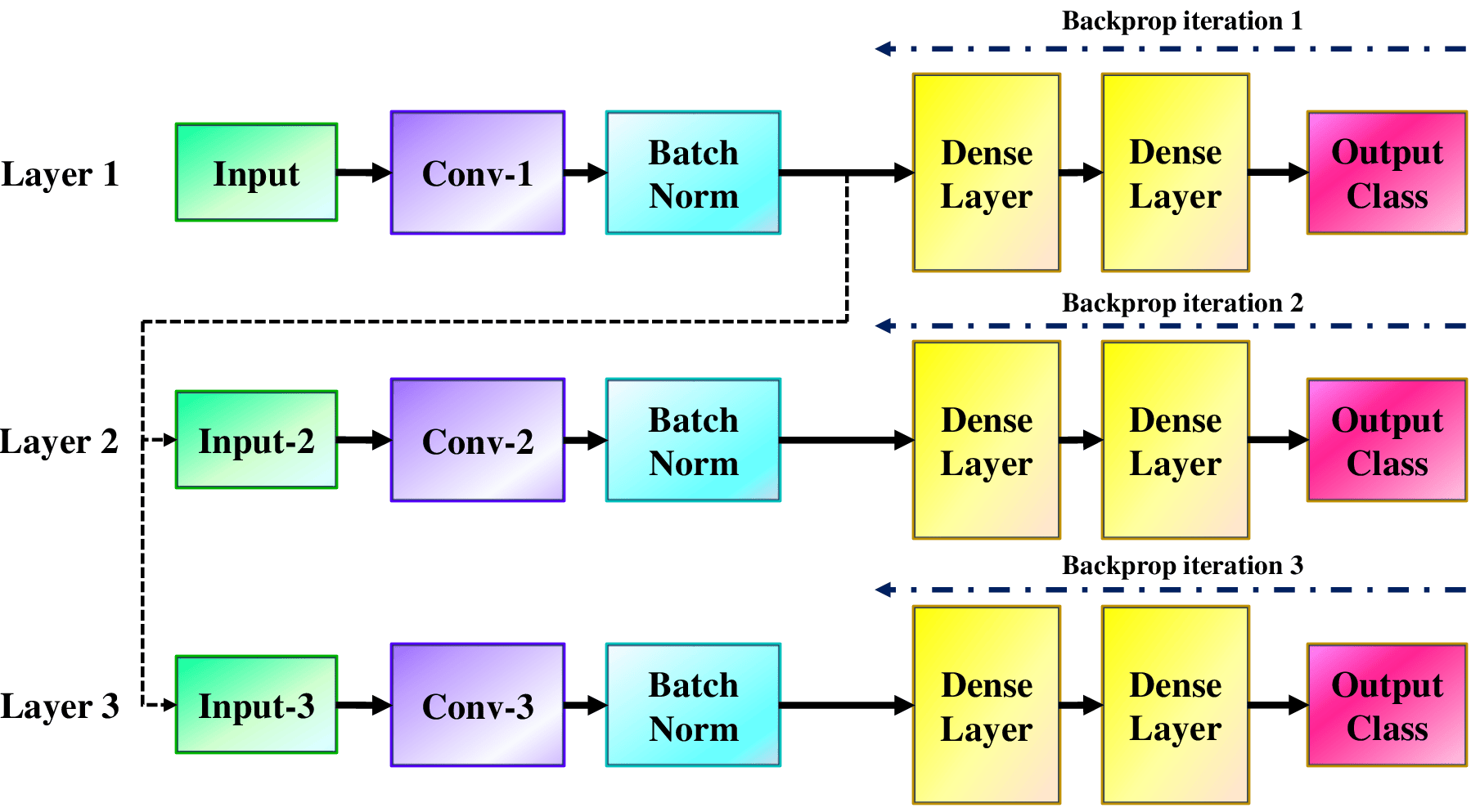}
    \caption{\small Deep Cascade Learning model} 
    \label{fig: dcl}
    \vspace{-4mm}
 \end{figure}

 Although the convolution network has demonstrated remarkable performance and results \cite{mukherjee2020energy}\cite{xing2022spectrum}%\cite{krichen2023convolutional}
 , it experiences the vanishing gradient problem while undergoing training \cite{noh2021performance}. This occurs because the weight updates during back-propagation are substantially reduced as the depth of the network increases, meaning that layers closer to input experience a slower learning rate \cite{marquez2018deep},\cite{ali2023scene}. 
 One solution to overcome this situation is the deep cascade learning algorithm proposed in \cite{marquez2018deep}. Deep cascade learning gradually trains the network from the lowest to higher layers. When the jamming signal is weak and therefore SJNR values are close to each other, the classifier faces trouble classifying between two classes. Deep cascade learning fine-tunes the weights by dividing the network into sub-layers and sequentially trains each layer until all the input layers have been trained. Through this technique, the vanishing gradient problem can be eliminated by compelling each layer of the network to acquire features that are correlated with the output. In other words, it maintains the linear correlation between input and output while accommodating the nonlinear relationship \cite{gautam2021comparative}. Furthermore, it has shown a significant reduction in training time and memory while adjusting the complexity of the network to the given data \cite{marquez2018deep}\cite{du2019transfer}. The structure of the DNN model using deep cascade learning is shown in Fig. \ref{fig: dcl}. The input layer is connected to the output using two dense layers. The weights for the initial model layer and the output are subsequently obtained via a back-propagation algorithm during the training. After reaching a state of stability, the second layer is trained by connecting to an output layer with a similar structure to the first layer using data created by forward propagating the actual inputs through the fixed initial layer. Through several iterations, every layer goes through this process for learning, and the weights remain constant for the subsequent layer. Adapting this approach helps increase the robustness of the second DNN and preserve the back-propagated gradient while having hidden layers. 
\graphicspath{{figures/}}
\vspace{-2mm}

\section{Simulation and Experimental Results}
\label{results}
%In this section, dataset generation and the performance of the proposed deep jamming detection architecture in offline training and online testing are evaluated and discussed thoroughly.
% \vspace{-4mm}
\subsection{Simulation results}
%__________Data Generation_________________
\subsubsection{Data generation}
To collect the received signal recordings, a synthetic 5G dataset is generated using MATLAB R2024a from the scenarios with and without a jammer under different conditions. The generated dataset is then processed by the data preprocessing block and fed into the detection model. A detailed description of the dataset generation environment is provided as follows.

The transmitter produces 5G waveform including four OFDM symbols containing the SS block. It is capable of transmitting the signal using four different types of modulation including QPSK, 16QAM, 64QAM, 256QAM. The signal is then passed through CDL-D channel model and is received by the UE. UE and gNB are positioned in different locations, and the UE is connected to the base stations with different cell IDs. Different locations for the UE results in different path loss experiences. The simulation environment is equipped with free-space path loss model to demonstrate the effect of different UE positions in the scenario as, \vspace{-2 mm}

\begin{equation}\label{eq:fspl}
    L_{fs}(dB) = 20\log \left(\frac{\lambda^2}{4\pi^2d^2}\right)
\end{equation}
\noindent where $\lambda$ and $d$ signify the wavelength and the distance between gNB and UE, respectively.
The received signal for each observation is recorded in the time domain, and the number of observations is $\mu$. The same process is performed to record the received signal in the presence of the jammer with different SJNR values and jamming powers. The number of observations in this scenario is denoted as $\mu_j$. The jammer used for the training is designed using AWGN. Other types of jammers, such as the jammer which can transmit modulated BPSK, and 8QAM signals, are used for testing. The dataset parameters used for the dataset generation are extracted from \cite{3gpp.38.901}, \cite{han2021hybrid} and listed in Table \ref{tab: sim param}. A total of $\mu=\mu_j=12{,}300$ observations are generated for each class of unjammed and jammed received signals using the parameters listed in Table \ref{tab: sim param}. These observations are evenly distributed across all modulation types and and three PSS indices of $N_{ID}^{(2)}\in\{0,1,2\}$.
\begin{table}[htbp]
\caption{Dataset Parameters} \vspace{-5 mm}
\begin{center}
\begin{tabular}{|c|c|}
\hline
Parameter & Value  \\
\hline\hline
SJNR & -10 to 30 dB, step size: 1 dB \\ \hline
Distance & 10 to 500 m, step size: 20 dB \\ \hline
%Dataset Size $N_{obs}$ & 18600 \\ \hline
Input Size & $5 \times 1024$ \\ \hline
$N_{FFT}$ & 2048\\ \hline
Delay Spread & 30 ns\\ \hline
Subcarrier Spacing (SCS) & 30 kHz\\ \hline
Sample Rate\tablefootnote{The sample rate equals to $SCS\times N_{FFT}$ which equals $61.44 $ Msps} &  61.44 Msps \\ \hline
Antenna Noise Temperature & 290 K\\ \hline
Cyclic Prefix & Normal \\ \hline
Number of Resource Blocks & 106 \\ \hline
Channel Model & CDL-D \\ \hline
gNB Power & 30 dB \\ \hline
Modulation Scheme & QPSK, 16QAM, 64QAM, 256QAM\\ \hline
\end{tabular}  
\end{center}
\label{tab: sim param}
 \vspace{-4mm}
\end{table}
%_______________DNN1_______________________
% training curve  
% ROC (10 dB - 20 dB SJNR)

\subsubsection{Performance under the Single Threshold DNN} \label{Sec: Single Threshold}
\begin{figure}[htp]
    \centering
    \includegraphics[width=0.8\linewidth, height=5cm]{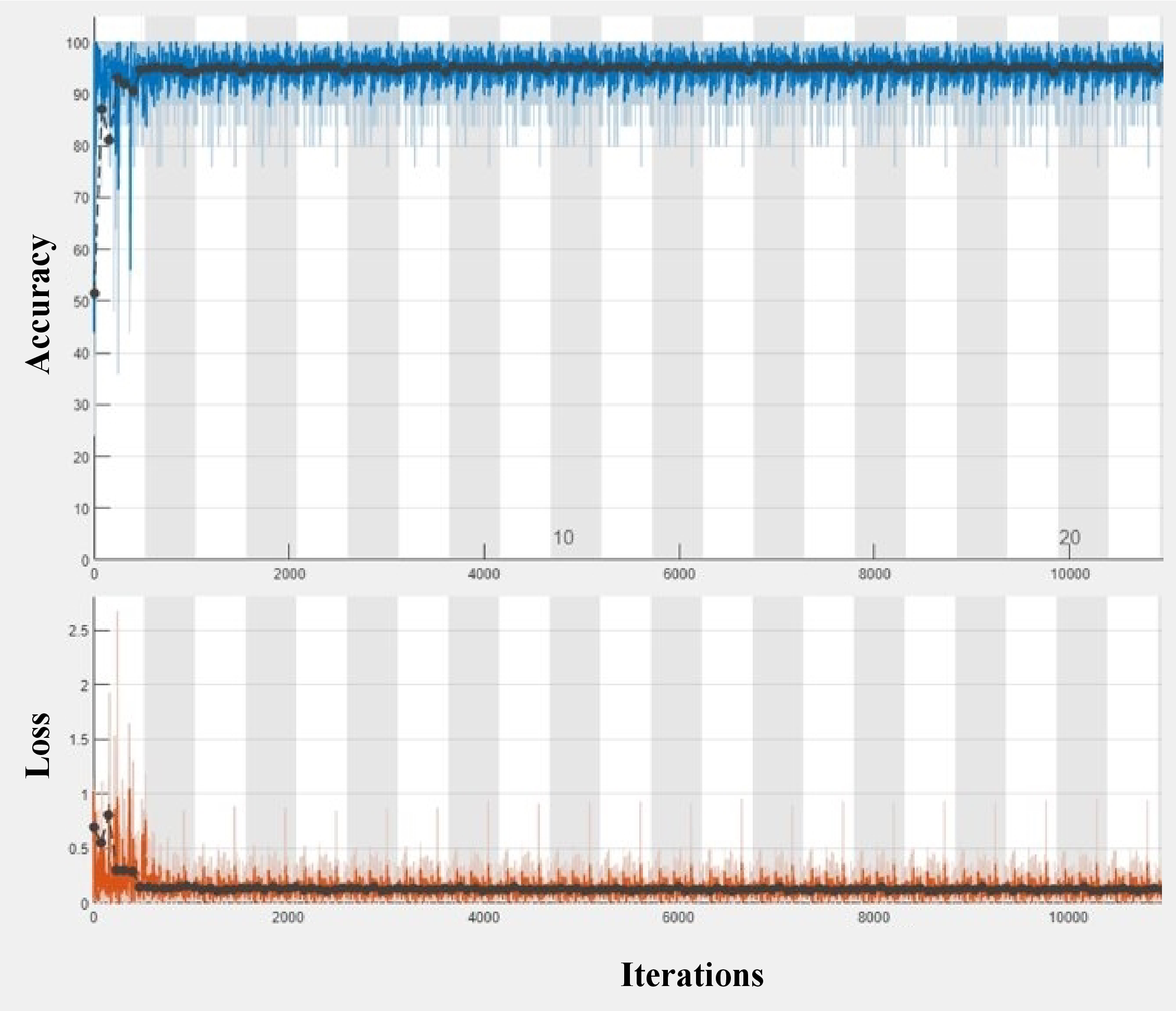}
    \caption{\small Offline training-- single threshold DNN model } 
    \label{fig: st-train} \vspace{-3 mm}
    %\vspace{-5mm}
 \end{figure}

The training performance of the first single threshold DNN is presented in Fig. \ref{fig: st-train}. In this figure, the upper graph plots the accuracy of the model which stabilizes around 94.93\% by jumping quickly into the convergence region in the early stages of the training process. The integration of DWT into the design results in more precision and less training time of 144 minutes, while before adding DWT the training time was around 41 hours. The below graph presents the loss plot which demonstrates a sharp drop during the early stages, eventually reaching a state of stability around 0.01. This signifies a decrease in the prediction error as the training proceeds. It can be concluded that the model is capable of classifying the unobserved data without falling into overfitting. 

\graphicspath{{figures/}}
\vspace{-0mm}

\begin{figure*}
    \centering
    \begin{subfigure}{0.23\textwidth}
        \includegraphics[width=1\textwidth, height=4cm]{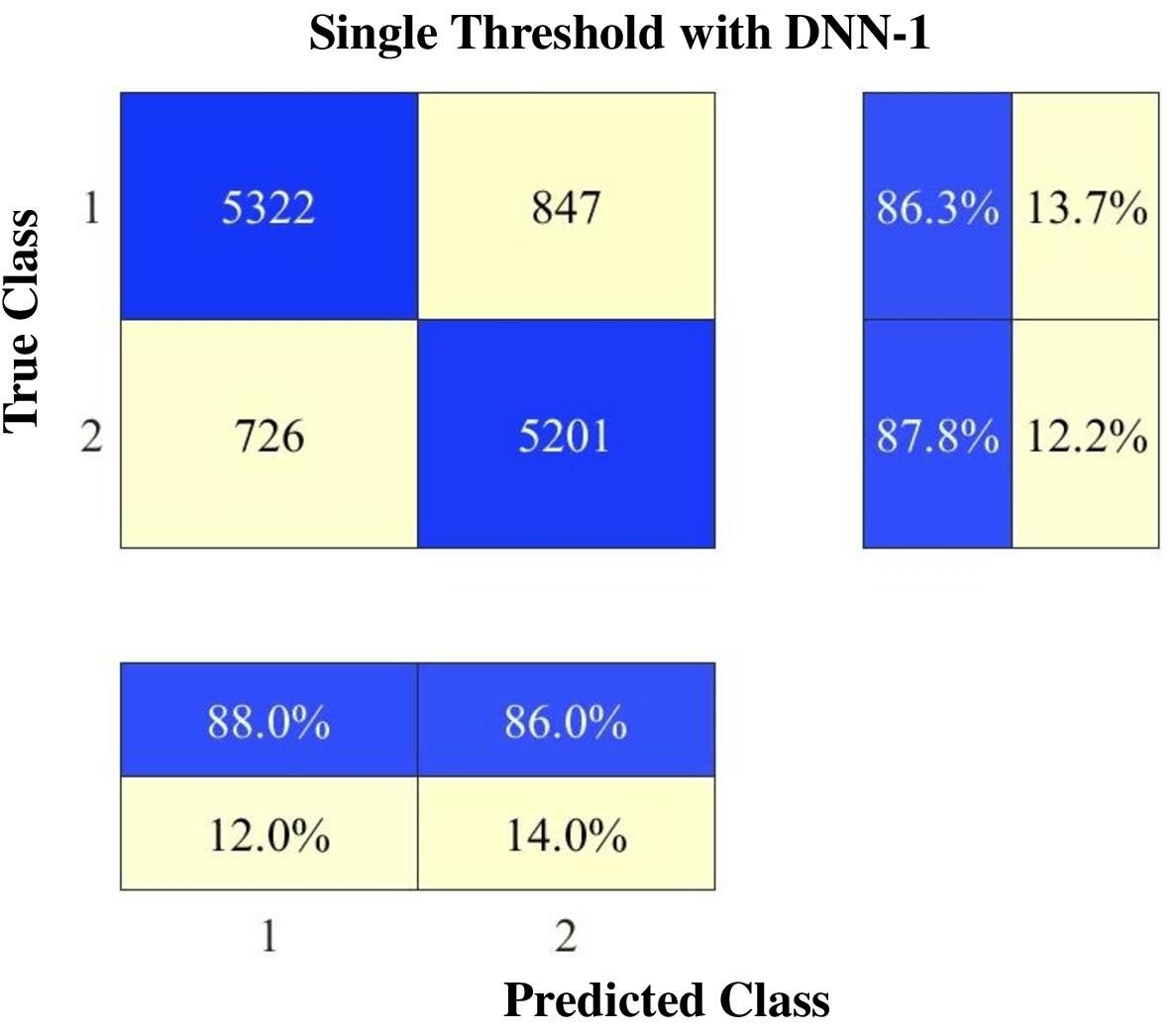}
        \caption{} 
        \label{fig: st-conf}
    \end{subfigure}
    \hfill
    \begin{subfigure}{0.23\textwidth}
        \centering
        \includegraphics[width=1\textwidth, height=4cm]{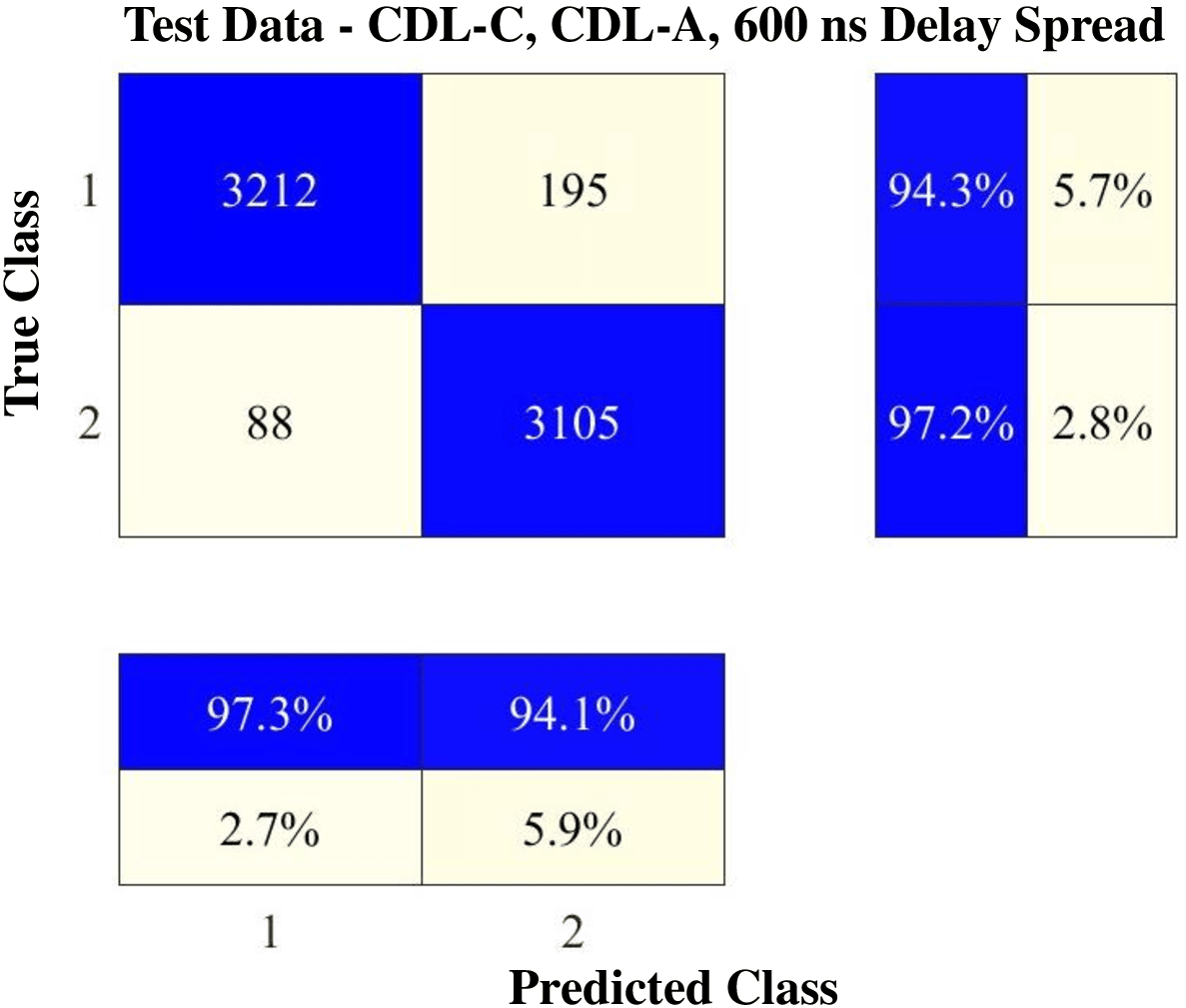}
        \caption{}
       \label{fig: channel-conf}
    \end{subfigure}
    \hfill
    \begin{subfigure}{0.23\textwidth}
        \includegraphics[width=1\textwidth, height=4cm]{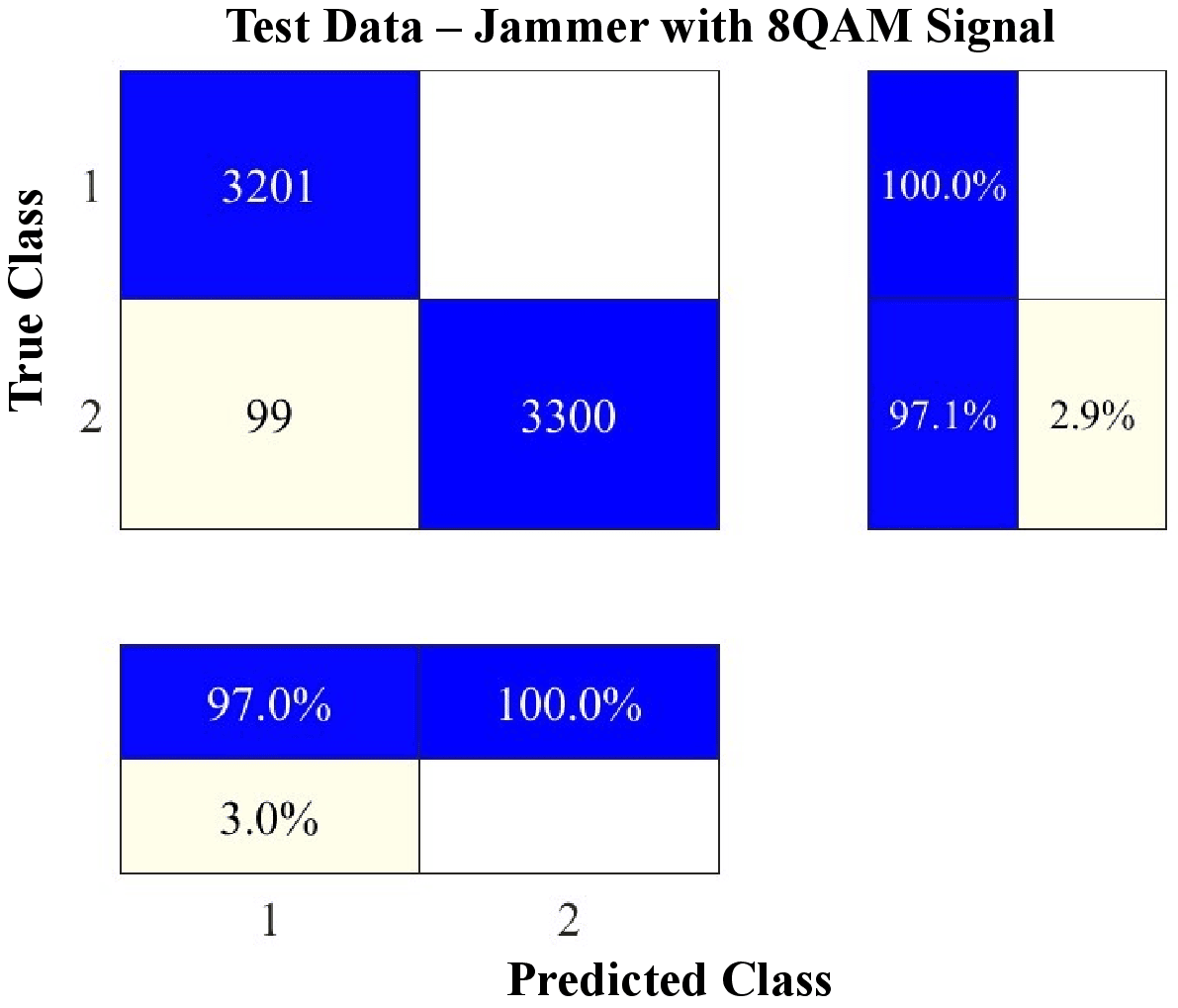}
        \caption{} 
        \label{fig: 8qam-conf}
    \end{subfigure}
    \hfill
   \begin{subfigure}{0.23\textwidth}
        \centering
        \includegraphics[width=1\textwidth, height=4.15cm]{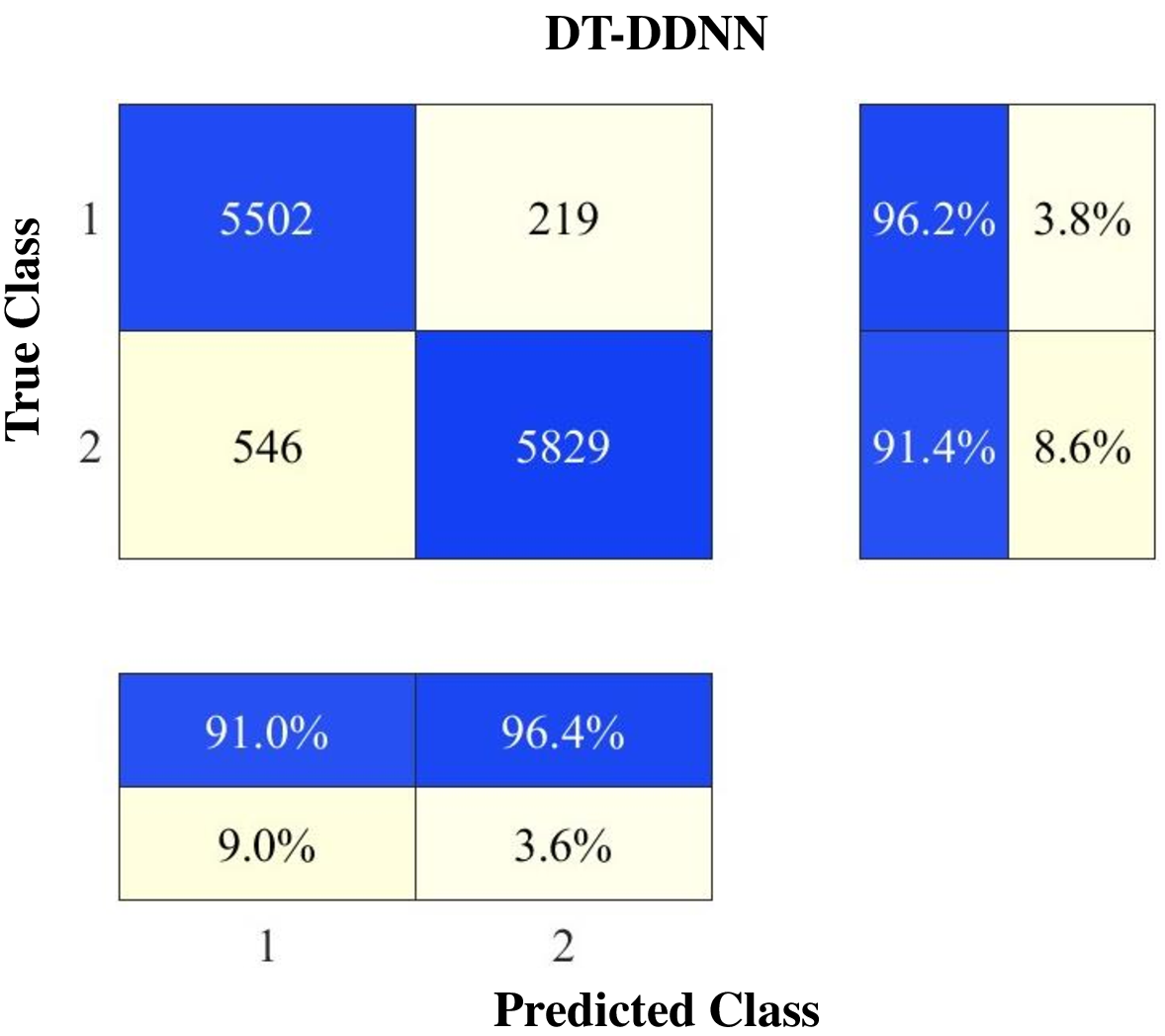}
        \caption{}
       \label{fig: dt-conf}
    \end{subfigure}

   \caption{(a) \small Confusion matrix of single threshold DNN-- SJNR from -10.5 to 30 dB. (b) Confusion matrix of single threshold DNN under test-- The test dataset consists of signals that have been transmitted through CDL-C and CDL-A channel models with a delay spread of $600ns$, the SJNR value is from 0 to 20 dB. (c) Confusion matrix of single threshold DNN under test-- a test dataset influenced by a jammer using 8QAM modulation. (d) Confusion matrix for the DT-DDNN with SJNR ranging from -10.5 to 30 dB.}
    \label{fig: conf-matrix}
    \vspace{-5mm}
\end{figure*}

Fig. \ref{fig: st-conf} presents the confusion matrix for jamming detection with a single threshold DNN design while the SJNR spans from -10.5 to 30 dB. The figure demonstrates the performance of the detection algorithm in which true positive (TP) and true negative (TN) correspond to non-jammed and jammed classes respectively. Based on this matrix, the algorithm classified 5322 observations as TP with the rate of 88.0\%, and 5201 cases as TN with the rate of 86.0\%. 12.0\% of the jammed cases were falsely classified as non-jammed scenarios and 14.0\% of the non-jammed observations were erroneously classified as jammed cases. This demonstrates the ability of the model to accurately differentiate between jammed and non-jammed signals. However, it also introduces the potential for further improvement to decrease the misclassification cases particularly to reduce the FN observations. In the jamming detection scenario, reducing FN is more important than FP, as failure to identify a jammer leads to security risks and communication disruption \cite{greco2021jamming}.

\graphicspath{{figs/}}

 % DNN1 tests
 % Channel, delay spread
 After training and testing the model, a different test dataset is implemented consisting of signals transmitted over CDL-C and CDL-A channel models with $600ns$ delay spread. The purpose of this test is to show the generalization of the proposed model to various 5G network settings. The confusion matrix in Fig. \ref{fig: channel-conf} demonstrates the robustness of the design to the disparities of the network by correctly detecting 97.3\% of the non-jammed cases and 94.1\% of jammed cases. Furthermore, recall and precision can be obtained as 0.97 and 0.94 respectively. Hence, the model demonstrates minimal misclassification which shows its ability to detect the attack despite increased complexity. This test highlights the potential of the detection model in a real-world 5G network while maintaining performance and accuracy in a more complex test environment.

% jamming type 8QAM
Besides testing under different network settings, a second test dataset compromised by a jammer transmitting with 8QAM signal modulation is created to test the effectiveness of the detector model to different types of jamming signals. Based on the confusion matrix of this test which is provided in Fig. \ref{fig: 8qam-conf}, despite using AWGN as the jamming signal in the training process, it demonstrates acceptable flexibility by accurately classifying 97\% of non-jammed cases and 100\% of jammed observations. These results show the robustness of the design against novel jamming techniques which signifies a transfer of learning from one jamming mode to another. However, no percentage of jammed cases were incorrectly classified, which corresponds to zero probability of miss-detection (FP=0). These results demonstrate the ability of the model to stay generalized to various types of jamming without requiring retraining.

%_______________DNN2_______________________
% training
% confusion matrix
% ROC/ total
% compare with DNN1
\subsubsection{DT-DDNN Performance}
\begin{figure}[htp]
    \centering
    \includegraphics[width=5.5cm, height=4cm]{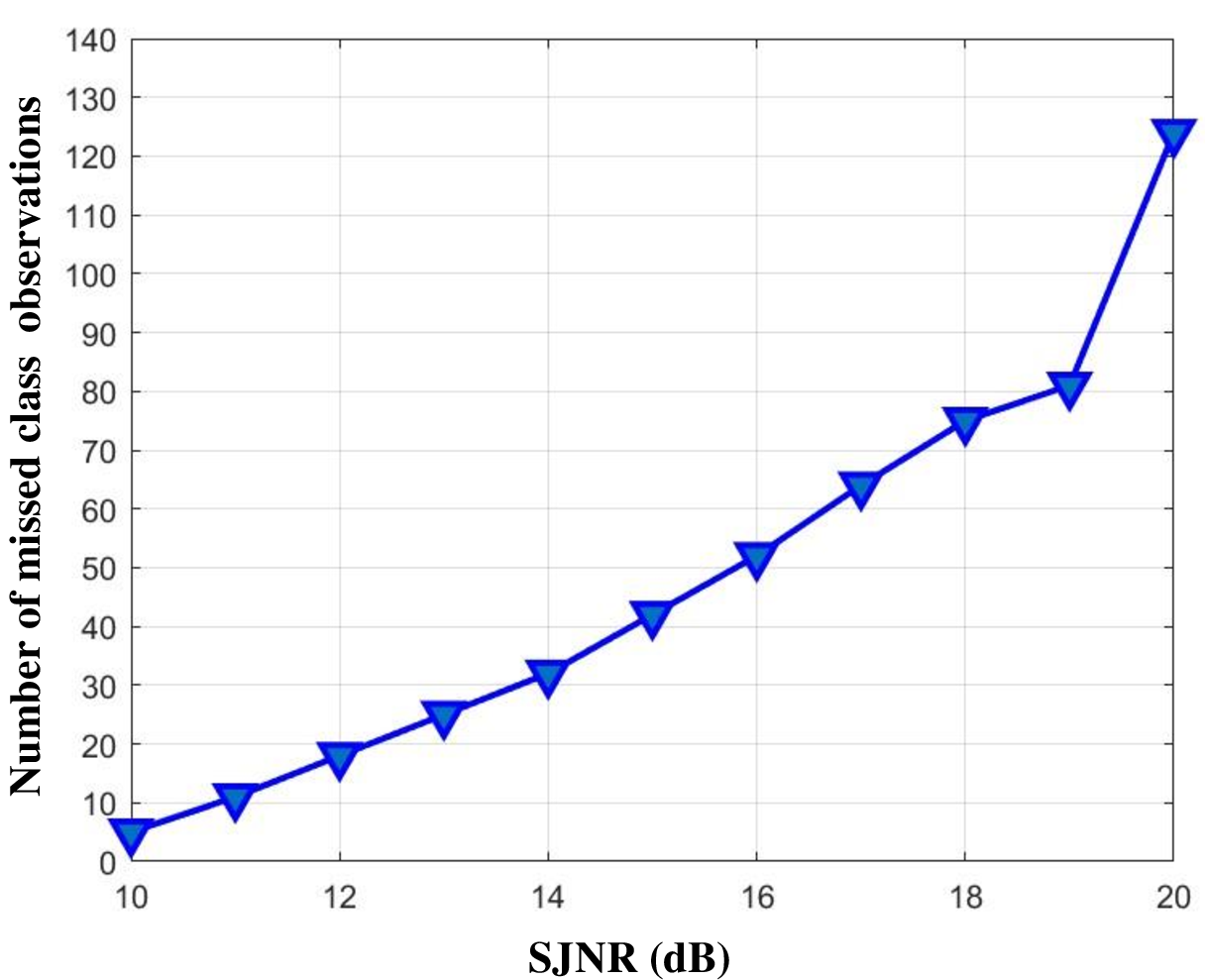}
    \caption{\small SJNR vs miss-classed observations} 
    \label{fig: SJNR} \vspace{-2 mm}
    %\vspace{-2mm}
 \end{figure}

Fig. \ref{fig: SJNR} demonstrates the relationship between the number of missed class observations and the SJNR in dB. During these experiments, the received power of gNB is fixed to focus solely on the effect of the jammer. It can be observed that as the SJNR increases, there is a significant rise in the number of missed detections. This trend is related to the reduction in the strength of the received jamming signal compared to the power received from the gNB as SJNR increases. As a result, the effect of the jammer on model features is decreased and the received signal pattern is similar to the cases without the jammer, which contributes to a rise in classification error. The observed situation highlights a significant challenge in the jamming detection model as it is more difficult for the detector to distinguish between the jammed and legitimate signals. Thus, in a high SJNR regime, the deep learning model incorrectly classifies jammed signals as normal. One solution to resolve this issue is adding a double-threshold DNN model to the design. The primary objective of this change is to augment the sensitivity of the jamming detection model, particularly in conditions characterized by high SJNR. By employing a double threshold approach, the detection model is capable of conducting a more comprehensive analysis of the signal, distinguishing low-power jamming attacks from regular signal fluctuations which enhances the precision of the model.
\graphicspath{{figures/}}
\vspace{-3mm}

The confusion matrix of the DT-DDNN design which includes a double DNN architecture to improve the sensitivity of the jamming detector is presented in Fig. \ref{fig: dt-conf}. While the first DNN uses a double-threshold concept to increase the sensitivity, the second DNN includes a deep cascade learning model to enhance the ability of the system to accurately classify complex observations and decrease the probability of miss-detection to 9.0\%. The model accurately classifies 91.0\% of non-jammed cases and 96.4\% of jammed observations, which shows almost 3\% and 10.4\% rise in the accuracy compared to the single threshold architecture. These results represent an advancement in the ability to detect and provide countermeasures for jamming attacks in real-world scenarios of the 5G network. 

The proposed method is compared to various state-of-the-art approaches in the literature. We tailored the proposed methods in \cite{viana2023deep} and \cite{Varotto2023} according to our scenario. Similar to \cite{Varotto2023}, we feed unprocessed data (I-Q samples) to the pre-trained CNN-based classifier. Instead of using AE trained with single class, we use CNN-based supervised learning as higher performance was observed in low jamming power. The work in \cite{viana2023deep} uses Deep Attention recognition mechanism (DatR) with an LSTM network for both LoS and NLoS cases with the input features of SINR and RSSI fed to the ML module. To improve the detection performance in NLoS where the received signal power is low, augmentation and majority voting are combined. Table \ref{table_comparison} compares accuracy, TP, and TN values for single-threshold, DT-DDNN, unprocessed I-Q \cite{Varotto2023}, and DAtR \cite{viana2023deep}. 
% For DAtR, two window sizes, $\omega=30$ and $\omega=100$, have been explored. 
For the DAtR model, the 1-D structure described in \cite{viana2023deep} is used, with two window sizes, $\omega=30$ and $\omega=100$, evaluated for comparison. Under test data with high SJNR, the DT-DDNN exhibits the accuracy, TP, and TN of 0.9368, 0.9144, and 0.9617, respectively, representing the highest performance. DAtR with $\omega=100$ and the proposed single threshold experience similar accuracy. While TP of the former is higher than the latter, DAtR yields lower TN than the single threshold method. 

%\vspace{-3.9mm}
\begin{table}[htbp]
\centering
	\caption{Comparison with other methods}
	\begin{tabular}{|c|c||c|c|} \hline
		\textbf{ Method}	&	Accuracy  & TP  & TN  \\
		\hline \hline
Single Threshold & 0.879 & 0.8775 & 0.8627 \\ \hline
\textbf{DT-DDNN} & \textbf{0.9368} & \textbf{0.9144} & \textbf{0.9617} \\ \hline
Unprocessed I-Q \cite{Varotto2023} & 0.8393 & 0.8501 & 0.8300 \\ \hline
DAtR \cite{viana2023deep} $\omega=30$ & 0.8540 & 0.8700 & 0.8403 \\ \hline
DAtR \cite{viana2023deep} $\omega=100$ & 0.8719 & 0.9001 & 0.8515 \\ \hline
\end{tabular}  
\label{table_comparison}
\vspace{-3mm}
\end{table}

\begin{figure}[htp]
    \centering
    \includegraphics[width=0.9\linewidth, height=6cm]{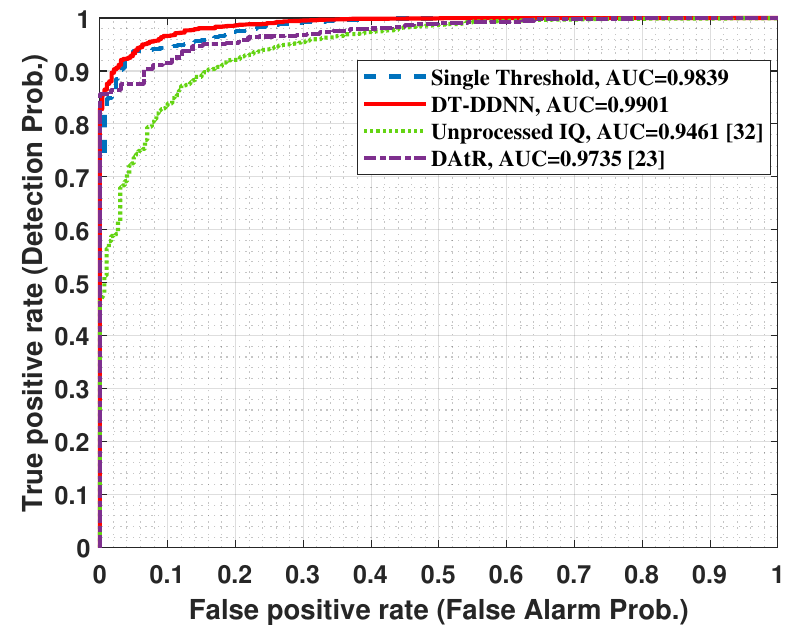}
    \caption{\small ROC curves comparing the DNN trained with unprocessed IQ samples and two proposed jamming detection designs.}
    \label{fig: dt-roc} \vspace{-6 mm}
    %\vspace{-2mm}
 \end{figure}

Comparative analysis of three DNN designs for jamming detection in a 5G network is provided by the ROC curves in Fig. \ref{fig: dt-roc}. The green dotted curve corresponds to a DNN trained on raw IQ samples taken from 5G waveform with no preprocessing block in the design. The optimum configuration was achieved with 4 layers of CNN with batch and ReLu following each layer. The single threshold DNN design, illustrated by the blue dashed curve, is correspondence to the single threshold DNN design that uses a single DNN to classify jammed cases. The ROC for DAtR is differentiated with dashed-line in purple in which $\omega=100$ is used due to improved performance. In contrast, the red solid curve represents the DT-DDNN design which uses a double DNN system to improve the sensitivity of the detection algorithm. The primary DNN block detects jamming by applying a double threshold decision-making method, and the secondary DNN handles observations that fall into the ambiguous area that the first DNN has found difficulties in classifying. By integrating a deep cascade learning model into the second DNN, the objective is to further enhance the classification performance. Based on the ROC curves, it is evident that the DT-DDNN provides a higher probability of detection ($P_D$) in comparison to the majority of false alarm probabilities ($P_{FA}$). This indicates a greater proportion of true positive and a reduced potential of miss-detection of legit signals as jamming. This suggests that the second design with DT-DDNN architecture offers a more robust jamming detection system under challenging detection cases. It is worth noting that since the received signal is affected by the communication channel, providing an ML module with at least one feature independent of the channel behavior, EPNRE herein, helps improve the detection performance.

\graphicspath{{figures/}}
\vspace{-5mm}

\subsection{Experimental Validation}
\begin{figure}[htp]
    \centering
    \includegraphics[width=6cm, height=5cm]{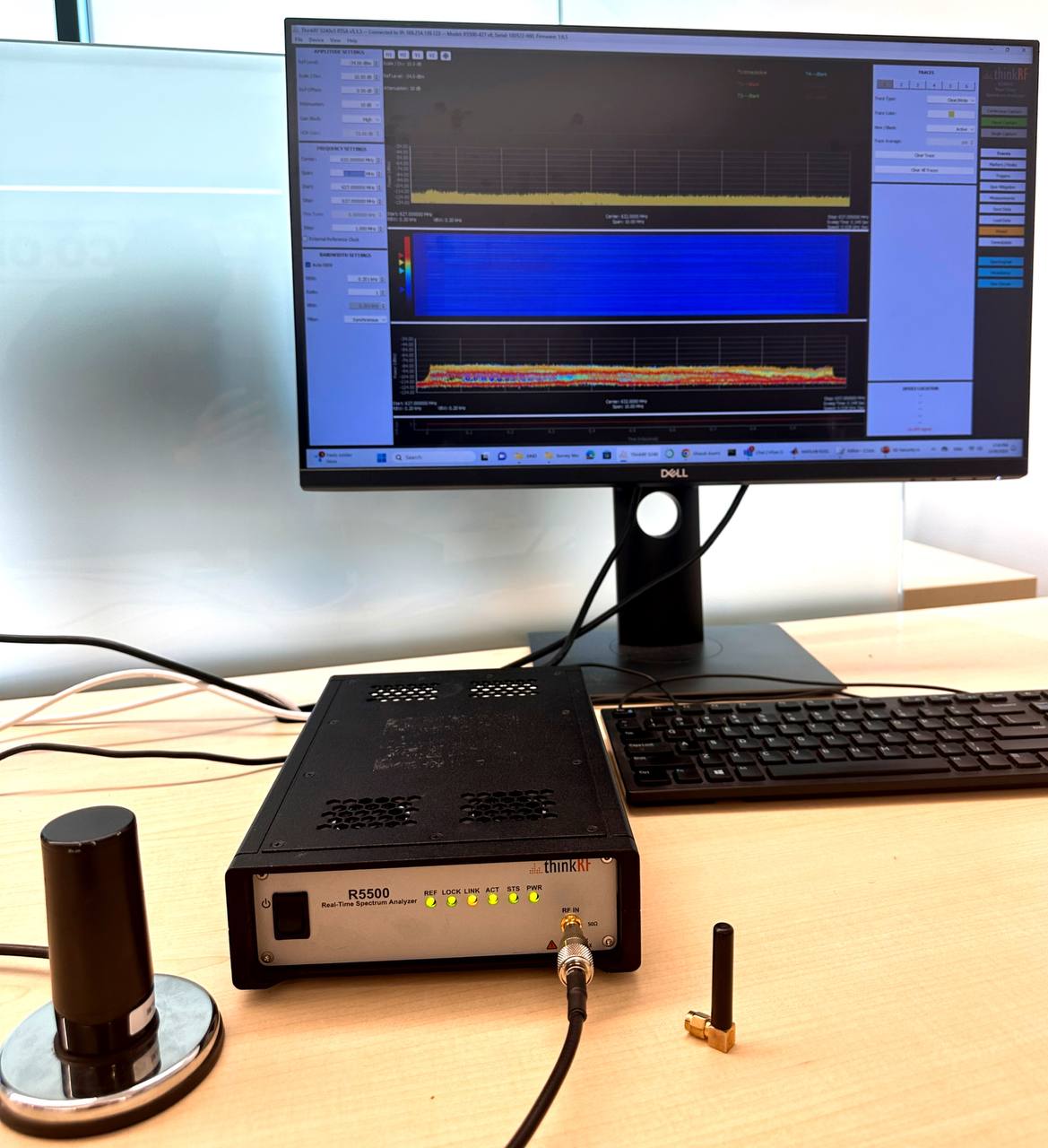}
    \caption{\small Experimental setup to collect real over-the-air dataset. The testbed includes \emph{thinkRF} RTSA R5500 spectrum analyzer, two types of antennas, and a PC. The spectrum analyzer represents a CAV receiver equipment which takes samples from the 5G RF domain and sends these samples to the Cloud in PC for preprocessing and DNN-based jamming detection.}
    \label{fig: equip} \vspace{-2 mm}
    %\vspace{-2mm}
 \end{figure}

% Experimental setup Description
The experimental evaluation is conducted within the FR1 5G NR n71 band operating within the downlink frequency range of 617-652 MHz and bandwidth of 35 MHz \cite{3gpp.38.104}. During the initial tests, it is concluded that this spectrum is shared between Telus (with carrier frequency of $f_c=632\ MHz$) and Rogers (with carrier frequency of $f_c=622\ MHz$). Data acquisition is performed using ThinkRF spectrum analyzer RTSA R5500 (shown in Fig. \ref{fig: equip}), which represents CAV receiver equipment, and two different types of antennas. The location of the test and therefore distance from the gNB is variable during the sampling. The experimental setup is configured with the sample rate of 15.625 MHz, the carrier frequency of $f_c=632\ MHz$, and the intermediate frequency bandwidth (IFBW) of 10\ MHZ. Sampling is conducted in various environments, including indoor (behind the windows and under the desk) and outdoor (Line-of-Sight (LOS) and Non-Line-of-Sight (NLOS)) scenarios and collected samples are stored in CSV format using PyRF4 API. 

\graphicspath{{figures/}}
\vspace{-0mm}

\begin{figure}[htp]
    \centering
    \includegraphics[width=0.8\linewidth, height=5 cm]{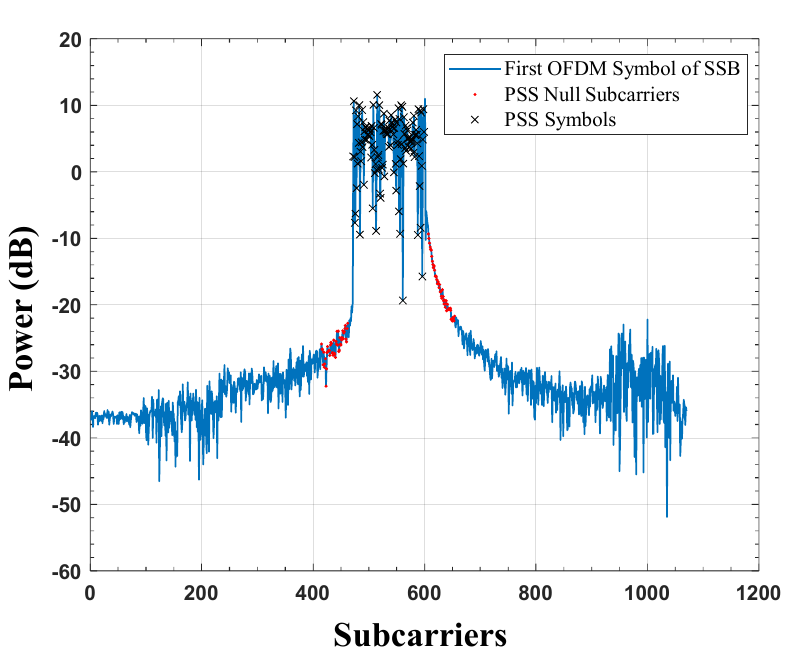}
    \caption{\small OFDM symbols of the extracted SSB based on subcarrier indices.}
    \label{fig: exp-pss} \vspace{-5 mm}
    %\vspace{-2mm}
 \end{figure}

% Synchronization with gNB and blind search
To extract precise information of SSB, it is critical to perform time offset (TO) and carrier frequency offset (CFO) estimations. This is due to the lack of knowledge of the exact center frequency, which necessitates the application of a blind search. To accurately calculate the TO and CFO, we utilize the PSS correlation characteristics along with the cyclic prefix from CP-OFDM 5G waveform to fine-tune with the gNB signal.
The optimization problem for estimating CFO is formulated as, \vspace{-3mm}
\begin{equation}\label{eq: optimization CFO}
\begin{split}
      \hat{f}_{CFO} &= \arg\max_{f_i} \Bigg[\sum_{\tau} y(\tau)e^{j2\pi \frac{f_i}{f_s}\tau}x^{ssb}_l(t-\tau)  \big|_{l=0} \Bigg],
\end{split}
\end{equation}
For obtaining time offset to the SSB, Schmidl \& Cox approach \cite{Schmidl1997} is adopted which exploits the cyclic prefix in 5G waveform. Hence, the following optimization problem is solved numerically.\vspace{-5mm}

\begin{equation}\label{eq: optimization TO}
\begin{split}
      \hat{t}_{off} &= \arg\max_{t} \, M(t)=\frac{|P(t)|^2}{R(t)^2} ,
\end{split}
\end{equation}
in which $P(t)$ and $R(t)$ are as, \vspace{-5 mm}

\begin{equation}\label{eq: TO-P}
\begin{split}
      P(t)=\sum_{m=0}^{L-1} y^*(t+m)y(t+m+L)  ,
\end{split}
\end{equation}
and \vspace{-5mm}

\begin{equation}\label{eq: TO-R}
\begin{split}
      R(t)=\sum_{m=0}^{L-1}|y(t+m+L)|^2 
\end{split}
\end{equation}
respectively.
Following the extraction of SSB, the correlation signals and EPNRE values are calculated. Fig. \ref{fig: exp-pss} demonstrates OFDM symbols of SSB versus the subcarrier indices for one of the observations selected as an example. In this figure, black cross markers represent the PSS symbols that are used to calculate PSS correlation, and red dots correspond to null subcarriers from which EPNRE is calculated.
The features of collected samples, including 6000 observations, are then transferred into a 3D tensor and fed into the DT-DDNN model. The threshold parameters of the DT-DDNN which defines the sensitivity of the model are updated based on environmental noise power.

% Extract energy and PSS
\graphicspath{{figures/}}
%\vspace{-1mm}

\begin{figure}[htp]
    \centering
    \includegraphics[width=5.5cm, height=5cm]{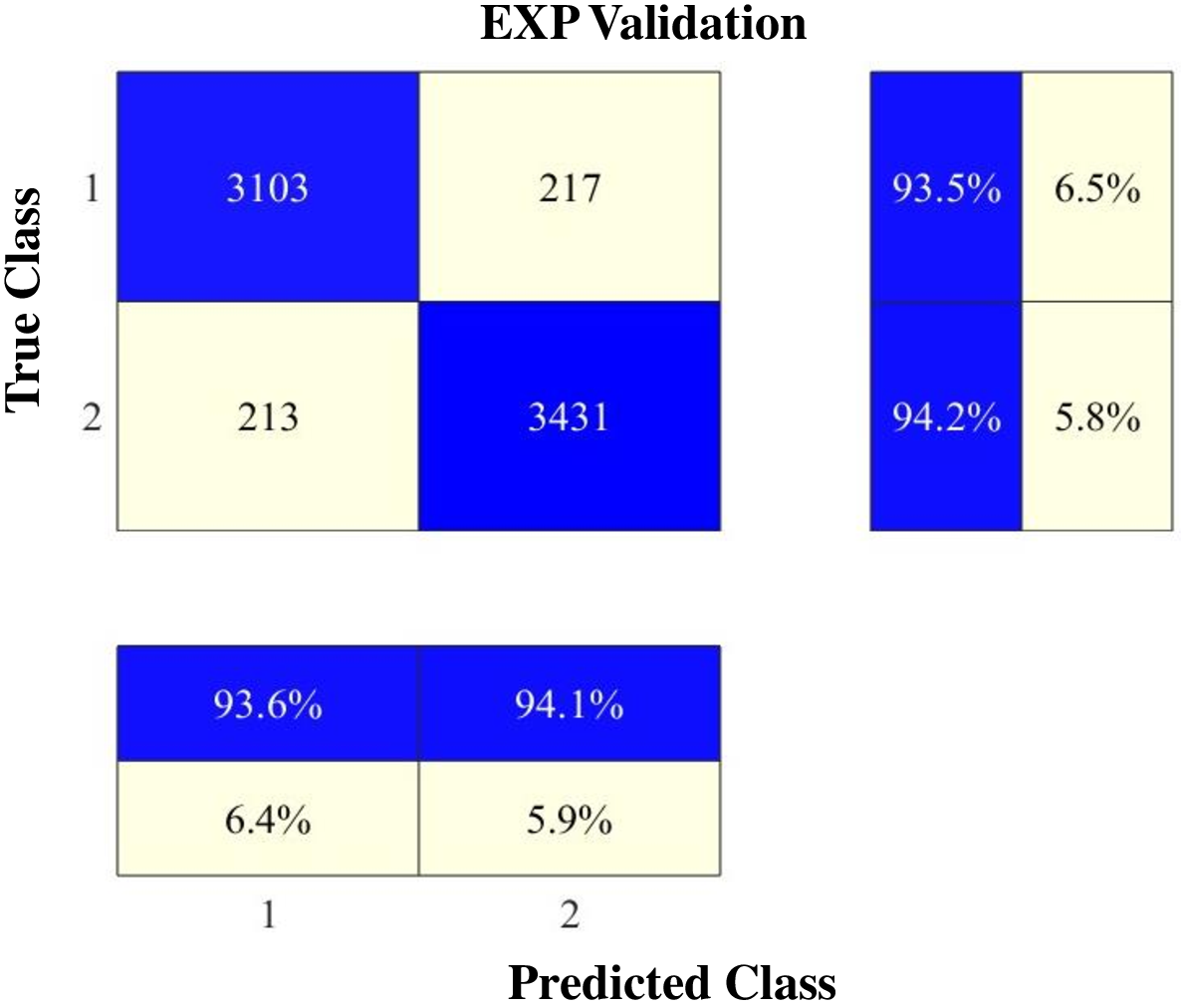}
    \caption{\small Confusion matrix of the performance of DT-DDNN in response to data obtained from the experimental setup.} 
    \label{fig: exp-conf} 
    \vspace{-2mm}
 \end{figure}

% results
The confusion matrix in Fig. \ref{fig: exp-conf} provides an evaluation of the performance of DT-DDNN model using data obtained from the practical testbed. Based on these results, the model accurately classifies 93.6\% of non-jammed cases and 94.1\% of jammed observations. 213 observations of jammed signals are classified as non-jammed cases with a miss-detection rate of 6.4\% and 217 cases of non-jammed observation are classified as jammed observations with false-alarm probability of 5.9\%. These results validate the practical applicability and efficacy of DT-DDNN model by showing the ability of the proposed model to accurately distinguish between jammed and non-jammed 5G signals in experimental configuration. 

\graphicspath{{figures/}}
\vspace{-2mm}

\section{conclusion}
\label{conclusion}
This work presents a robust deep-learning-based approach to detect smart and barrage jamming attacks in 5G networks with a particular focus on 5G SSB. By incorporating deep cascade learning and DWT, our DT-DDNN architecture provides remarkable accuracy in classifying a wide range of jamming scenarios, including those characterized by high SJNR values and diverse signal transmission settings. A preprocessing block is integrated to extract the PSS correlation and EPNRE characteristics of the received signal which has enhanced the ability of the model to differentiate between the jammed and non-jammed observations. With the inclusion of a DWT block in the model, the performance of the training process has enhanced and the training duration is reduced while maintaining the accuracy of the detection. Our results show that DT-DDNN outperforms the single threshold approach and provides more robustness and sensitivity to different jamming scenarios. DT-DDNN demonstrates improved detection probability by 10.4\% and 13.2\% compared to single threshold design and unprocessed IQ sample DNN design respectively. Furthermore, the adaptability of the model has been verified via several experiments with varied channel conditions, delay spread, and jamming techniques. The provided outcomes highlight the effectiveness of the suggested approach which precisely identifies jammer presence in the network with minimal false positive and miss-detection. Additionally, an experimental setup is built to assess the performance of the proposed DT-DDNN model in response to real 5G signals. The experiments conducted on the data collected from the testbed confirm the effectiveness of the system in practical scenarios.
Our ongoing research includes an investigation of the potential of unsupervised generative models in jamming detection to improve the performance of the system in the detection of unseen and novel jamming attacks. %Furthermore, combining the CFO estimator with the design is expected to transfer the preprocessing load partially to the machine learning block.

\section*{Acknowledgment}
This work was supported in part by funding from the Innovation for Defence Excellence and Security (IDEaS) program from the Department of National Defence (DND), and in part by the Natural Sciences and Engineering Research Council of Canada (NSERC) CREATE TRAVERSAL Program. The authors would like to acknowledge the support of \emph{thinkRF} in building the experimental setup for this work.
\bibliographystyle{IEEEtran}
%\bibliography{references}

\vspace{12pt}

\end{document}